\documentclass[journal]{IEEEtran}
\usepackage[utf8]{inputenc}
\usepackage{cite}

\ifCLASSINFOpdf
   \usepackage[pdftex]{graphicx}
   \graphicspath{{../figures}{../bio}}
   \DeclareGraphicsExtensions{.pdf,.jpeg,.png,.eps,.tikz,.jpg}
\else
\fi
\usepackage[cmex10]{amsmath,empheq}
\usepackage{array}

\ifCLASSOPTIONcompsoc
 \usepackage[caption=false,font=normalsize,labelfont=sf,textfont=sf]{subfig}
\else
 \usepackage[caption=false,font=footnotesize]{subfig}
\fi

\usepackage{stfloats}
\usepackage{url}

\usepackage[dvipsnames]{xcolor}
\usepackage{flafter}

\usepackage{multirow}
\usepackage{colortbl}

\begin{document}
%
\title{Modelling of Supercapacitor Banks for Power System Dynamics Studies}
%
%
%

\author{Matej~Krpan,~\IEEEmembership{Student~Member,~IEEE,}
        Igor~Kuzle,~\IEEEmembership{Senior~Member,~IEEE,}
        Ana~Radovanović,~\IEEEmembership{Student~Member,~IEEE,}
        Jovica V. Milanović,~\IEEEmembership{Fellow,~IEEE}
\thanks{M. Krpan and I. Kuzle are with the Department
of Energy and Power Systems, Faculty of Electrical Engineering and Computing, University of Zagreb, 10000 Zagreb, Croatia, e-mail: matej.krpan@fer.hr; igor.kuzle@fer.hr}
\thanks{A. Radovanović and J. V. Milanović are with the School of Electrical and Electronic Engineering, University of Manchester, Manchester, UK, e-mail: ana.radovanovic@manchester.ac.uk; milanovic@manchester.ac.uk}
\thanks{This work was supported by the EU Horizon 2020 project “CROSSBOW”, grant agreement No. 773430.}
}

\maketitle

\begin{abstract}
The paper presents accurate and simple dynamic model of a supercapacitor bank system for power system dynamics studies. The proposed model is derived from a detailed RC circuit representation. Furthermore, a complete control system of the supercapacitor bank is also presented. The proposed model is easy to integrate in any power system simulation software and consists of only up to four standard datasheet parameters. The performance of the proposed model in grid frequency control and low-voltage ride through is illustrated on IEEE 14-bus test system in DIgSILENT PowerFactory. It is shown that in case of transient stability simulations the ideal (simplified) model of the supercapacitor can be used while in case of frequency control the ideal representation may not always be appropriate.
\end{abstract}


%
\IEEEpeerreviewmaketitle

\bstctlcite{bstctrl}

\section{Introduction}
%
%
%
%
\IEEEPARstart{T}{he} trend of increasing inverter-interfaced generation (IIG) in power systems throughout the world and subsequent reduction of synchronous inertia has motivated many research efforts to understanding stability of low-inertia systems as well as developing new algorithms which enable the IIG participation in system frequency control and other ancillary services. 

Supercapacitor (SC) energy storage system (ESS) can be used both as a standalone ESS for grid support or as a combination with other storage systems or IIG as part of a hybrid energy (storage) system. Its high power density, in particular, as well as hundreds of thousands of charging/discharging cycles and fast discharge naturally make it most applicable during grid frequency excursions when a fast injection or absorption of active power is necessary. Similarly, it can be used to quickly stabilize intermittent output of solar and wind generation. There are several reasons for using SC systems for fast injection of high power instead of other storage devices, e.g. batteries or flywheels \cite{Luo2015, maxwell2018}: i) SC bank can be fully charged or discharged in the time scale of several tens of seconds or faster while the rated power can be reached within a few milliseconds; ii) SCs have bigger power density than batteries and flywheels, i.e. a SC system of the same power rating will be much smaller than equivalent battery or flywheel system; iii) SCs can withstand significantly more (hundreds of thousands) charging/discharging cycles; iv) SCs have smaller operation and maintenance costs than batteries and flywheels.

\subsection{Literature survey}\label{sec:literature}
SC technology has often been used for electric vehicle applications in the past, e.g., \cite{Tahri2018}, with the focus on numerical modelling and/or energy management. SCs are often paired with wind and/or solar generation systems for power smoothing, virtual inertial response or low-voltage ride through (LVRT)\cite{Zhou2014, Arani2013, Gkavanoudis2014, Fang2018} and in these papers the SC is usually used in the DC link of voltage source converters. On larger scale applications, SC is often used as a part of a hybrid ESS in microgrids, e.g. \cite{Fini2018}, or isolated power systems \cite{Gevorgian2017, Sigrist2015, Cao2018, Molina2009, Dellile2012, Egido2015} for leveling out intermittent renewables or for grid ancillary services such as frequency or voltage support.

All of the surveyed papers have one or more similarities: i) SC is modelled as an ideal capacitor which is not always appropriate as the capacitance, and therefore the stored energy as well, varies with the applied voltage \cite{Musolino2013};  ii) SC energy storage system model is not applicable for power system dynamic studies as the supercapacitor model is either given in its RC/RLC form (not a block diagram with defined inputs and outputs), e.g. \cite{Molina2009, Rafik2007} or the complete control system is only given functionally on a higher level with actual details on subsystem implementation missing which makes it difficult to integrate in power system simulation software, e.g. \cite{Dellile2012}. One outlier from the reviewed literature is a simplified model by Egido \textit{et al.} \cite{Egido2015} which neglects any capacitor dynamics and is described only with an initial state-of-charge (SoC) estimation and a simplified control system which seems to agree fairly well with field measurements. However, the used disturbances were smaller than the size of a fully charged supercapacitor and the time scale was not long enough to observe the differences in time-to-discharge (i.e. when the stored energy is depleted). As will be shown in this paper, both initial SoC and the size of a disturbance impact the performance of different supercapacitor models.

To the best of our knowledge, there were no detailed studies on the adequate complexity level of a SC model for power system dynamics studies, no rigorous derivation of the SC bank model and no complete representation of a SC bank system with associated control that can be easily integrated in commercially available power system simulation software. Hence, this paper is trying to bridge the identified gap in modeling.

\subsection{Contribution}\label{sec:contribution}
A complete supercapacitor/ultracapacitor bank model should include: accurate dynamics of a SC cell, SC DC current calculation, charge/discharge control, active power and voltage/reactive power inverter control as well as frequency control loop. Therefore, the contributions of this paper are as follows:
i) detailed analysis and comparison of different SC models with varying levels of detail; ii) development of the appropriate models of different components of SC based energy storage system for power system dynamics applications (voltage and frequency control, transient stability studies); iii) development of accurate dynamic model of the SC bank with all the necessary controls that is easy to integrate in any commercial power system simulation software.

\section{Derivation of Supercapacitor Bank Model}\label{sec:derivation}
\subsection{Supercapacitor theory}
\begin{figure}[!t]
\centering
\includegraphics[width=2.4in]{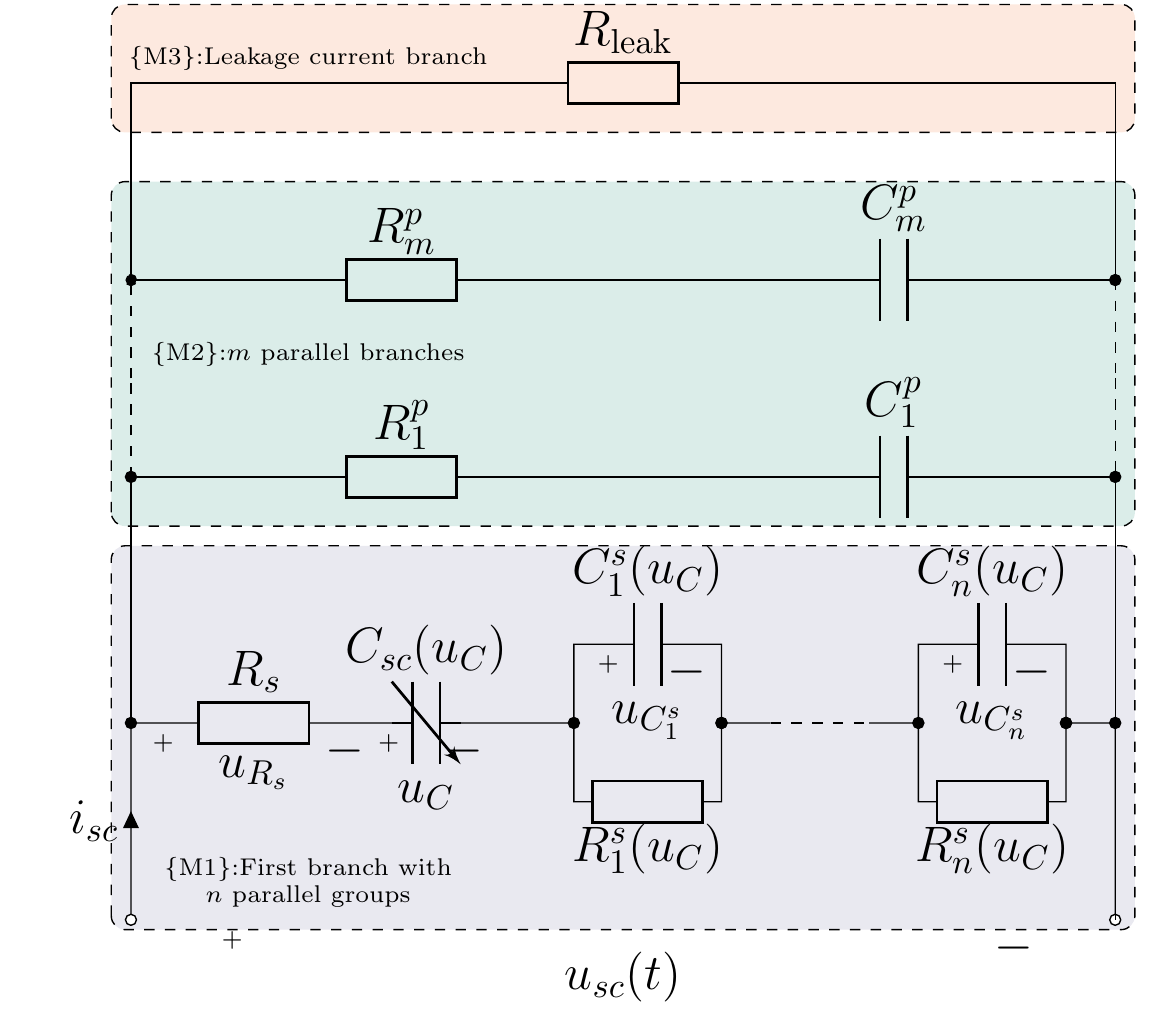}
\caption{Detailed RC circuit of a supercapacitor cell}
\label{fig:rc}
\end{figure}

Core of the SC bank model is the SC cell. Overview of different SC models can be found in \cite{Faranda2007} while the state-of-the-art SC models are available in \cite{Buller2002, Faranda2007, Musolino2013}. Basically, all these models are based on RC circuit identification using impedance spectroscopy. They can be described with the same type of RC circuit consisting of three parallel sections as shown in Fig. \ref{fig:rc}. The first branch $\{$M1$\}$ models fast dynamics, parallel branches $\{$M2$\}$ model slower recombination phenomena after a fast charge or discharge and the third branch $\{$M3$\}$ models the long-term self-discharge phenomena \cite{Musolino2013}. 

The time constants of the RC circuits in the parallel branches (ranging from a minute or several minutes up to an arbitrary amount of time as reported in \cite{Musolino2013, Faranda2007}) are much longer than the timescale investigated in this paper (up to around 30 s), hence they can be neglected in SC models developed here as will be shown in section \ref{sec:simplification}.

The important characteristics of SCs for model development can be summarized as follows: i) majority of the ultracapacitor capacitance comes from $C_{sc}$; ii) series combination of parallel branches $R_1^sC_1^s$--$R_n^sC_n^s$ is actually an infinite series of these parallel groups. However, it has been shown in \cite{Musolino2013} that five groups are sufficient to obtain an accurate model; iii) capacitance $C_{sc}$ as well as infinite sum elements $R_k^s$, $C_k^s$ are dependent on the ultracapacitor voltage $u_{C}(t)$. This dependence is nonlinear hence the model has the time-varying parameters. That is why an ideal capacitor representation used in many papers in the past is not always appropriate; iv) the number of parallel branches in the $\{$M2$\}$ group is   also theoretically infinite, though it has been shown that  between two and four branches are sufficient to achieve accurate results \cite{Musolino2013}.

In order to simplify the model further the following assumptions are made: i) $R_s$ is the equivalent series, voltage dependent resistance (ESR) determined at very high frequency \cite{Musolino2013}. Considering that $R_s$ is small ($<10$ m$\Omega$) and its impact on the model performance is insignificant, it is  typically considered to be constant; ii) temperature dependence of the parameters is neglected, i.e., the temperature is considered to be constant. The assumption is that the cooling of the system is adequate and that the system operates at room temperature. Although this effect can be included in the model, it was considered that for the initial derivation of the model for power system dynamic studies the temperature can be considered to be constant.

Parameters of the first branch are calculated according to (\ref{eq:uc1})--(\ref{eq:uc3}) \cite{Musolino2013}.
\begin{equation}\label{eq:uc1}
    C_{sc}(u_{C}) = C_0 + k_vu_{C}(t)
\end{equation}
\begin{equation}\label{eq:uc2}
    C_k^s = \dfrac{1}{2}C_{sc}, \quad k\in \{1...n\}
\end{equation}
\begin{equation}\label{eq:uc3}
    R_k^s = \dfrac{2\tau(u_{C})}{k^2\pi^2C_{sc}}
\end{equation}
$C_0$ is the ultracapacitor capacitance at 0 V and $k_v$ is a constant expressed in F/V. $\tau(u_C)$ is another experimentally determined parameter (it has a dimension of time) that can be expressed as function changing linearly with the voltage $u_C$: $\tau(u_{C}) = \tau_0 + k_\tau u_{C}(t)$ \cite{Musolino2013}. However, it can also be approximated by (\ref{eq:uc4}) \cite{Musolino2013}:
\begin{equation}\label{eq:uc4}
    \tau(u_{C}) \approx 3C_{sc}(R_{dc} - R_s),
\end{equation}
where $R_{dc}$ is the resistance experimentally obtained at very low frequencies (essentially DC). Naturally,  $R_{dc} > R_{s}$.

All the parameters of the $\{$M1$\}$ branch can be identified using manufacturer’s data sheet. 

Parameters of the branches {M2} and {M3} though, are more difficult to obtain since they must be obtained experimentally. Furthermore, these parameters are not universal and they depend on the time scale of the phenomena to be observed (described by the RC time constant $\tau_{RC} = RC$). The time scales are arbitrary, however they usually imply a range from several minutes to several weeks or even more \cite{Faranda2007, Musolino2013}. 

\subsection{Simplification of the supercapacitor cell model}\label{sec:simplification}
This section shows that the SC model for power system dynamics studies can be simplified and described by only the $\{${M1}$\}$ branch and with at least one parallel group  $R_k^sC_k^s$. The experimental data from \cite{Faranda2007, Musolino2013} are used to develop dedicated models in Matlab/Simulink and then use them in simulations for model simplification. The simulation results are related to two commercial SCs (Epcos 110 F and Maxwell 140) represented by different models with varying levels of detail to validate model simplification.

\begin{figure}[!t]
\centering
\subfloat[]{\includegraphics[width=2.1in]{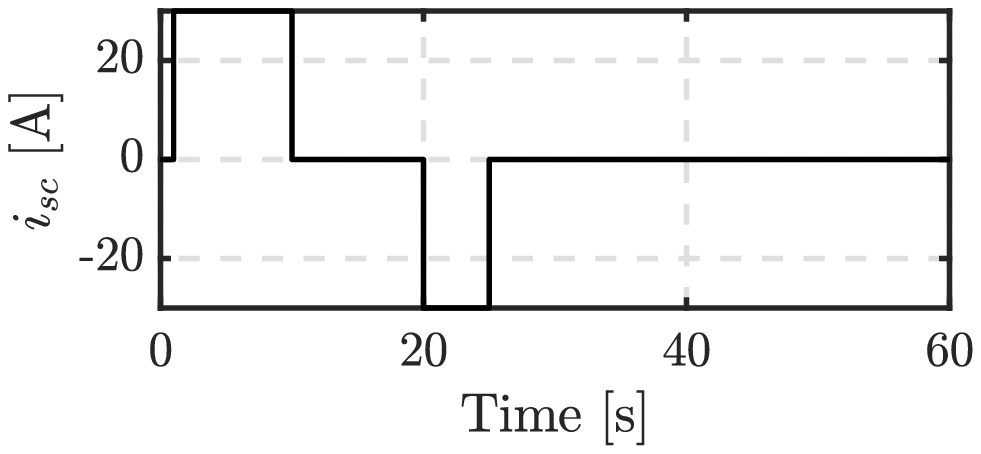}%
\label{fig:current}}
\hfil
\subfloat[]{\includegraphics[width=2.1in]{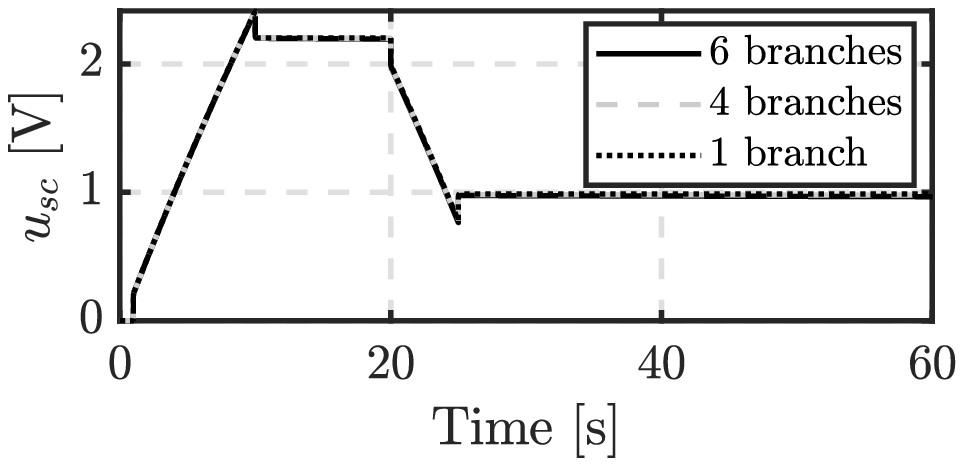}%
\label{fig:maxwell1}}
\hfil
\subfloat[]{\includegraphics[width=2.1in]{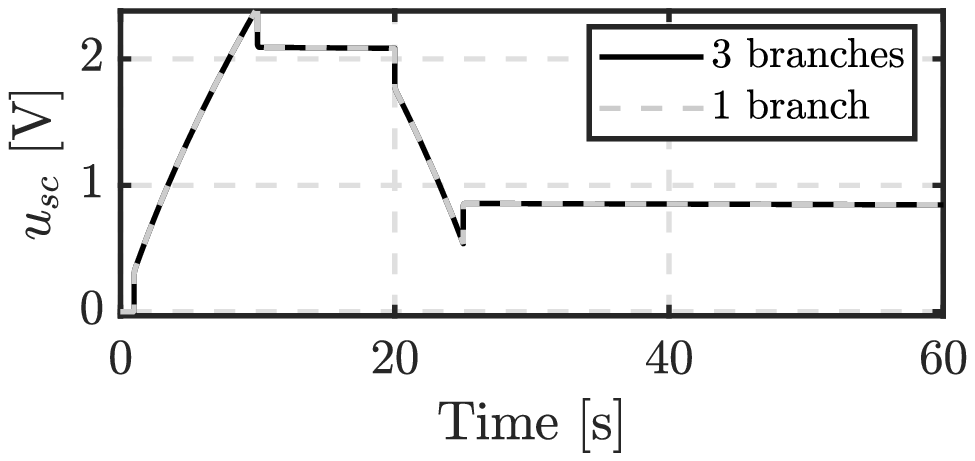}%
\label{fig:epcos1}}
\caption{Comparison of model response for different number of branches: a) Test current 30 A, b) Maxwell BCAP0140 voltage; c) Epcos 110 F voltage}
\label{fig:paralleltest}
\end{figure}

Different model responses are produced in MATLAB-Simulink using \textit{Simscape Electrical} toolbox. The number of branches in $\{${M2}$\}$ and $\{${M3}$\}$ groups is being sequentially reduced and the different model responses to the charge/discharge test are compared. Input to the model is the current $i_{sc}(t)$ and output of the model is the SC terminal voltage $u_{sc}(t)$. Results are shown in Fig. \ref{fig:paralleltest}. For clarification, 6 branch model represents the total number of branches (first branch $\{$M1$\}$, 4 parallel branches $\{$M2$\}$ and a self-discharge branch $\{$M3$\}$). Results for both SCs with different levels of detail (Fig. \ref{fig:maxwell1} and Fig. \ref{fig:epcos1}) show that the branches $\{$M2$\}$ and $\{$M3$\}$ do not have an impact on the model accuracy for the time scale of interest. Therefore, all the branches except the first branch can be neglected.

\begin{figure}[!t]
\centering
\subfloat[]{\includegraphics[width=2.1in]{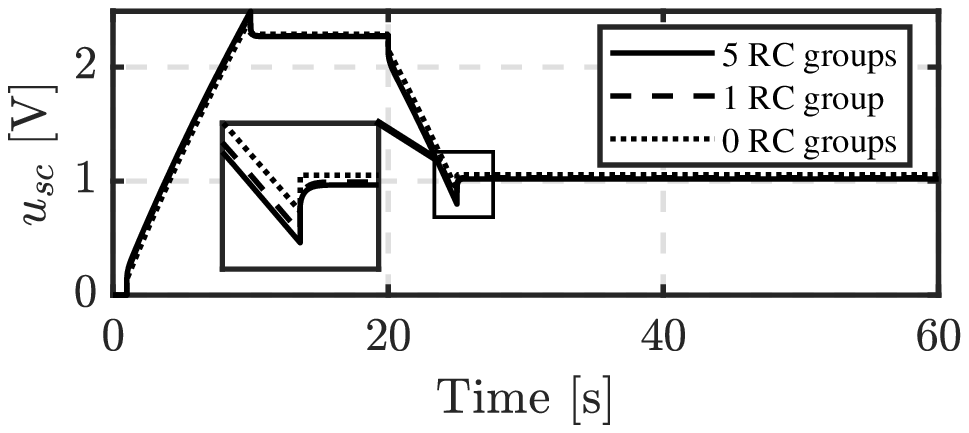}%
\label{fig:maxwell2}}
\hfil
\subfloat[]{\includegraphics[width=2.1in]{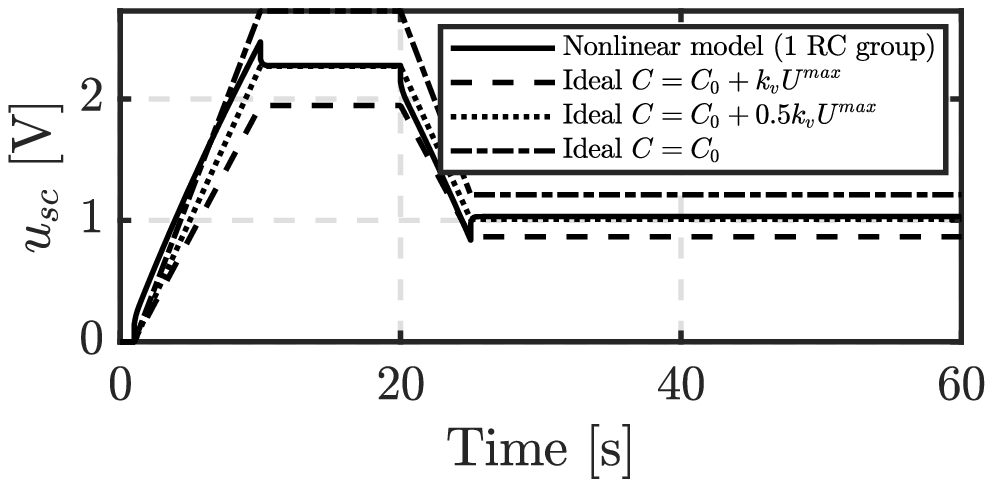}%
\label{fig:idealC}}
\hfil
\subfloat[]{\includegraphics[width=2.1in]{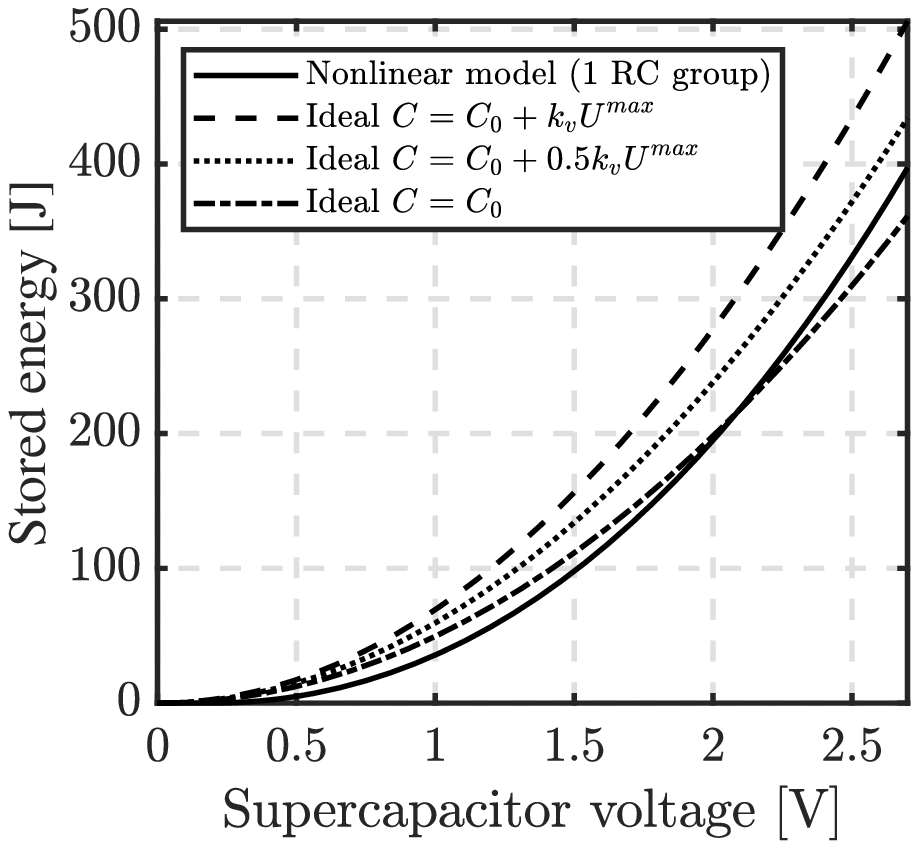}%
\label{fig:energy}}
\caption{Maxwel BCAP0140 model response: a) Voltage response to different numbers of RC groups; b) Voltage response compared to ideal capacitor; c) Energy stored in a supercapacitor with respect to voltage}
\label{fig:grouptest}
\end{figure}

In the next step, the adequate number of parallel RC groups in the first branch is determined. Results obtained for Maxwell 140 model are shown in Fig \ref{fig:maxwell2}. The significance of parallel RC groups depends on the difference between $R_{dc}$ and $R_s$. If those resistances are very close together (e.g. $<1 \text{m}\Omega$ as in Epcos model) then the impact on the accuracy is negligible. However, if the difference between $R_{dc}$ and $R_s$ is larger (e.g. $>2 \text{m}\Omega$ as in Maxwell 140 model) then the accuracy is significantly impacted which is important from both the available energy perspective and control perspective. Mean absolute percentage error for the Maxwell model with different numbers of groups is shown in Table \ref{tab:errorRC}. Since $R_{dc}$ and $R_s$ in reality depend on the cell in question, it can be conservatively concluded that at least one parallel group should be included: both $R_s$ and the parallel RC groups will cause voltage drops and energy losses in the circuit, thus should be included. Since the only voltage that can be measured is the one across the supercapacitor terminals ($u_{sc}$), $R_s$ and parallel RC groups will impact both the estimation of SoC and the DC voltage dynamics which determine the operating range of the supercapacitor module.

\begin{table}[!t]
 \renewcommand{\arraystretch}{1.3}
\caption{Accuracy of the supercapacitor model (Maxwell) with different numbers of parallel RC groups in the first branch}
\label{tab:errorRC}
\centering
\tiny
\begin{tabular}{|c||c|c|}
\hline
\multirow{2}{*}{Number of RC groups} & \multicolumn{2}{c|}{Mean absolute percentage error [\%]} \\
\cline{2-3}
& Voltage & Energy\\
\hline
1  (relative to 5 group model) & 2.6 & 6.5\\
\hline
0 (relative to 5 group model) & 6.7 & 12.9\\
\hline
0 (relative to 1 group model) & 4.3 & 10.3\\
\hline
\end{tabular}
\end{table}

Fig. \ref{fig:idealC} compares voltage response between the ideal capacitor and the nonlinear model with 1 RC group. The closest voltage profile was obtained when the capacitance of the ideal capacitor was set to the SC capacitance at half the rated voltage. Nevertheless, the ideal representation will not reflect the voltage transient effect due to the ESR which occurs when the charging or discharging current is discontinued. This voltage transient is important from the control perspective because it impacts the logic that enables or disables the charging/discharging based on state-of-voltage. More importantly, Fig. \ref{fig:energy} shows the difference between stored energy in a nonlinear model and in the ideal capacitor. Using capacitance at rated voltage is overly optimistic regarding the stored energy for the same applied voltage, while using capacitance at 0 V is closest to the nonlinear model (mean absolute percentage error in energy for the operating points between 0.5 V and 2.5 V is 27\%). This overestimation of the charge of the ideal ultracapacitor is coming from i) Different initial stored energy in the steady-state (capacitance of the ideal capacitor is constant while real capacitor has constant and variable, i.e., proportional to voltage, component); ii) Different rate of discharge because of the variable capacitance; iii) Internal losses due to equivalent series resistance and the parallel RC groups of the first branch in Fig. \ref{fig:rc}.

\subsection{Building a supercapacitor bank model}
In section \ref{sec:simplification} it was shown that the SC dynamics can be accurately represented using first branch only $\{$M1$\}$ with at least one parallel RC group. To build a capacitor bank of a higher power rating, a certain number of cells $n_s$ can be connected in series to form a string and a certain number of strings $n_p$ can be connected in parallel to form a module. Modules can then be connected in parallel to form a bank. Assuming completely identical cells, it is easily shown using Kirchoff's voltage and current laws that the voltage of the string $u_{sc}^s$ and the current of the module $i_{sc}^m$ are equal to (\ref{eq:vstring}) and (\ref{eq:imodule}), respectively.
\begin{equation}\label{eq:vstring}
    u_{sc}^s(t) = n_su_{sc}(t)
\end{equation}
\begin{equation}\label{eq:imodule}
    i_{sc}^m(t) = n_pi_{sc}(t)
\end{equation}

Assumption of identical and balanced cells is reasonable for bulk power system simulations, otherwise the modelling would have to be done on an electronic component level. Commercial solutions always have some sort of cell balancing system implemented to keep the cell usage and aging uniform. Cell aging is reflected in the change of parameters so that can be accounted for if the parameters of an older system are known.

\begin{figure}[!t]
\centering
\includegraphics[width=3.4in]{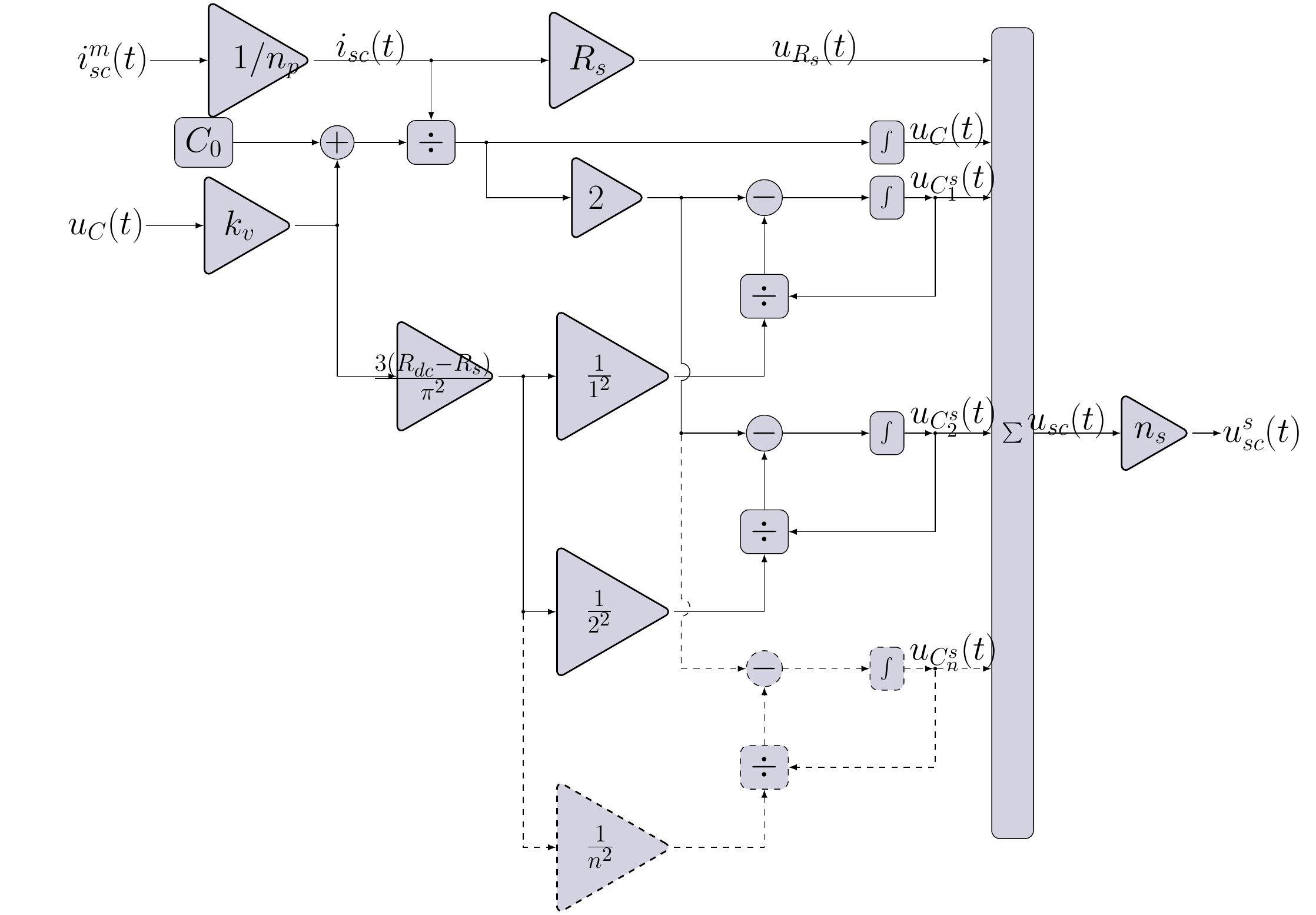}
\caption{Block diagram of the nonlinear supercapacitor module model}
\label{fig:scblock}
\end{figure}

Finally, the dynamic model of the bank can be built using circuit analysis in the time domain for the first branch only by setting $u_{sc}(t)$ as an output $y(t)$, $i_{sc}(t)$ as an input $u(t)$. Capacitor voltages are chosen as state variables. Complete nonlinear model of the SC bank in the analytic form is described by (\ref{eq:y})--(\ref{eq:state}) and shown in Fig. \ref{fig:scblock} where $R_k^s$ and $C_k^s$ are defined by (\ref{eq:uc2}) and (\ref{eq:uc3}), respectively.

\begin{IEEEeqnarray}{c}
    u_{sc}(t) =  i_{sc}(t)R_s + u_C(t) + \sum_{k=1}^n u_{C_k^s} = y(t) \label{eq:y} \\
    i_{sc}(t) = u(t)\\
    u_{sc}^s(t) = n_s u_{sc}(t) = n_s y(t)\\
    i_{sc}^m(t) = n_p i_{sc}(t) = n_p u(t)\\
    \dfrac{du_C}{dt} =  \dfrac{i_{sc}(t)}{C_0 + k_v u_C(t)} \label{eq:state0} \\
    \dfrac{du_{C_k^s}}{dt} = -\dfrac{u_{C_k^s}}{R_k^sC_k^s} +  \dfrac{i_{sc}(t)}{C_k^s} \label{eq:state}
\end{IEEEeqnarray}

Note that (\ref{eq:state0})--(\ref{eq:state}) assumes the relation $dQ = C_{sc}\cdot du_C$ per \cite{Faranda2007} instead of $Q=C_{sc}\cdot u_C$. This is an alternative definition of capacitance for voltage-dependent capacitors \cite{Zeltser2018}. Nevertheless, in both definitions the expression for capacitance in the denominator of (\ref{eq:state0})--(\ref{eq:state}) remains linearly proportional to voltage only differing in the magnitude of $k_v$. The behavior remains the same and the capacitance is varied by varying $k_v$ in the case studies presented here.

SC cells are very low voltage devices (rated voltage up to 2.5--3 V) and they have to be scaled up to form a bank of grid-scale power rating, e.g. 1 MW to 100 MW. The number of cells necessary to form a grid-scale SC bank is in the order of $10^3$ for 1 MW bank up to $10^5$ for 100 MW scale.



\section{Supercapacitor control system}\label{sec:system}
\begin{figure}[!t]
\centering
\includegraphics[width=2.5in]{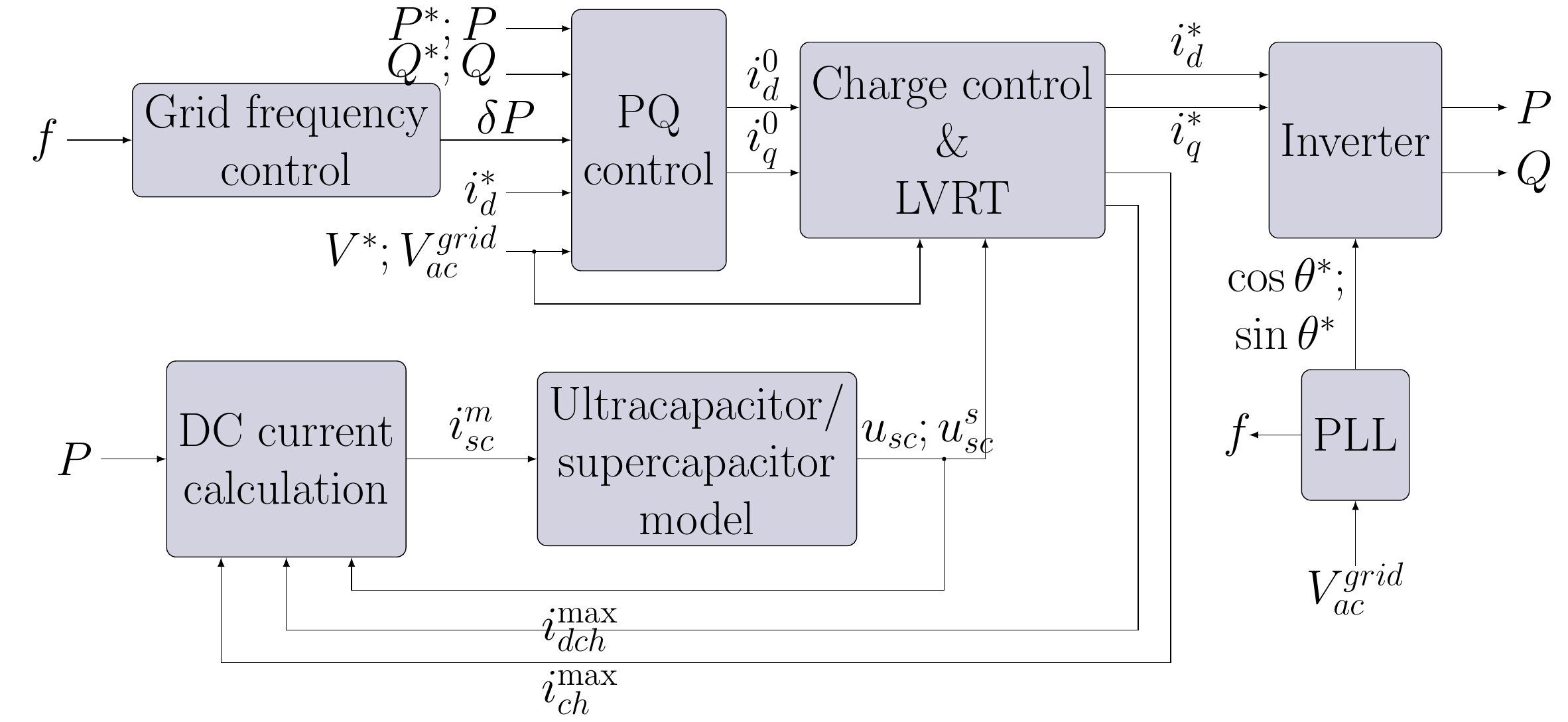}
\caption{Complete model of the supercapacitor bank system}
\label{fig:scsystem}
\end{figure}

The complete control system consists of inverter PQ control, charge/discharge control, DC current calculation and frequency control loop. The block diagram of the complete SC bank energy storage system is shown in Fig. \ref{fig:scsystem}. $P$ and $Q$ are active and reactive power injected or absorbed by the inverter to or from the grid, while asterisk (*) denotes a set-point value. $V_{ac}^{grid}$ is the AC voltage of the bus the inverter is connected to. $i_d$ and $i_q$ are the direct and quadrature axis currents of the inverter. Inverter is controlled in the grid voltage reference frame in which the grid voltage phase is estimated by the PLL (model version 3 in PowerFactory). PLL also estimates the system frequency for frequency control block. DC current calculation block calculates the SC current for charging or discharging. It should be noted that the presented SC control system is similar to a battery control system \cite{dspfbattery} since the requirements are the same (constant power output). The difference from the other control schemes however, is that the voltage measurement is directly used as a measure of energy (State-of-Voltage, SoV) rather than State-of-Charge (SoC) through current integration since the energy of a capacitor is directly proportional to the voltage. Furthermore, capacitors are much more sensitive to applied voltage, which varies significantly more than in batteries. Therefore, a special care must be taken not to overcharge the SC since even the voltage which is only 5\% above the rated voltage can damage the cell. Similarly the SC should not overly discharge because of current limitation for constant power. The individual controller blocks are further elaborated in the following subsections.

\subsection{Charge control}
\begin{figure}[!t]
\centering
\includegraphics[width=2.6in]{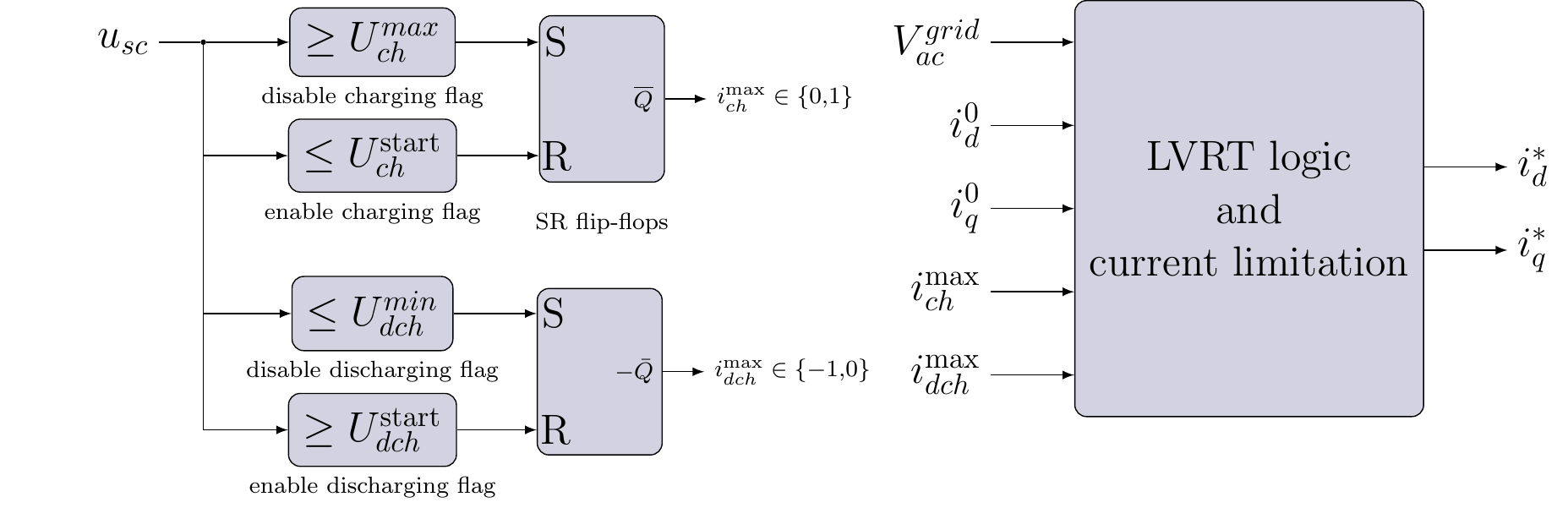}
\caption{Charge control, LVRT and current limitation block}
\label{fig:chcontrol}
\end{figure}

Fig. \ref{fig:chcontrol} shows the structure of this block. State-of-Voltage (SoV) measurement is used to control the charging and discharging process. Charging is stopped if the SC bank is charged to nominal voltage, while discharging is stopped when the SC voltage falls below a user defined low voltage threshold. Charging/discharging is enabled again when the voltage reaches a user defined minimum voltage level for charging/discharging. The input to the block are the $d$ and $q$ axis currents $i_d^0$ and $i_q^0$ from the PQ control, while the final inverter current set-points $i_d^*$ and $i_q^*$ are determined by this block. Simple low-voltage ride through logic and current limitation are also implemented in this block. They are not shown here since they are available in many papers, e.g. \cite{Gkavanoudis2014, dspfbattery, WECCguidelines}.

\subsection{DC current calculation}
\begin{figure}[!t]
\centering
\includegraphics[width=2.4in]{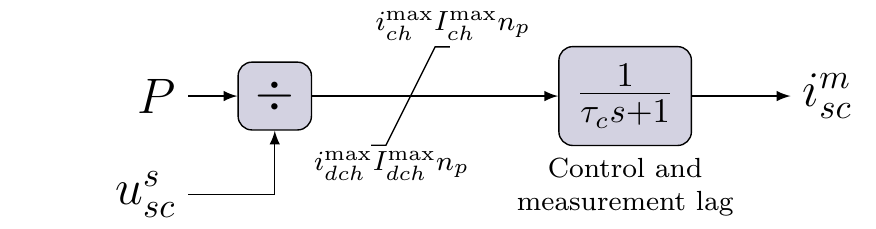}
\caption{DC current calculation block}
\label{fig:dc}
\end{figure}

The input to the SC model is the current, however in power system applications the power is usually controlled and not the current. This block calculates the charging or discharging DC current based on the actual inverter power output. Block diagram of this subsystem is shown in Fig. \ref{fig:dc}. It should be noted that this module as well as the SC model works with SI units, while other subsystems work in p.u. $I_{ch}^{\text{max}}$ and $I_{dch}^{\text{max}}$ are the maximum single cell charging and discharging current in A (e.g. $\pm 100$ A).

\subsection{PQ control}
\begin{figure}[!t]
\centering
\includegraphics[width=2.4in]{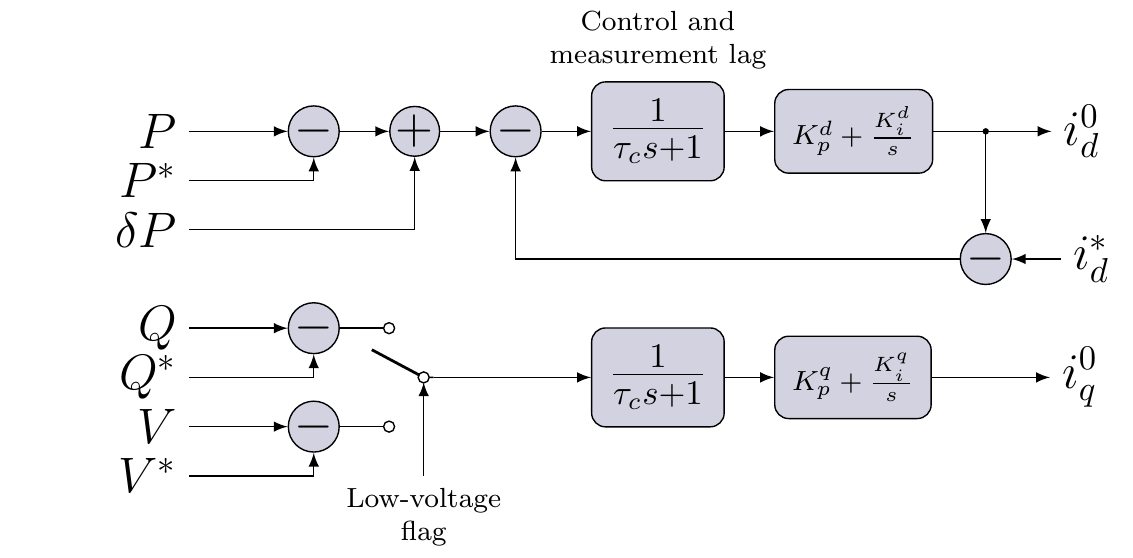}
\caption{Supercapacitor bank inverter PQ control}
\label{fig:pq}
\end{figure}

Fig. \ref{fig:pq} shows the PQ control structure of the SC bank inverter. In this case, the inverter is modelled as a controlled current source and the $d$ and $q$ axis currents are obtained from the active and reactive power control error, respectively. Measurement/control lag is also included in this block diagram. The term $i_d^* - i_d^0$ is a compensation term for active power during low-voltage ride through when the active power should be low and reactive power high. Reactive power or terminal voltage control can be both chosen. However, if reactive power control is chosen it will be overridden by terminal voltage control during low-voltage ride through. Simplified structure of the converter and its control applicable for electromechanical transient simulations of large networks (integration step size 1--10 ms) which neglects fast inner current control loops and AC-side filter/grid dynamics was used in this study \cite{WECCguidelines}. Other or more detailed models could be used for different studies, e.g. EMT, single-machine infinite bus or smaller systems \cite{Molina2009, Fang2018, Gevorgian2017}.

\subsection{Grid frequency control}
\begin{figure}[!t]
\centering
\includegraphics[width=2.2in]{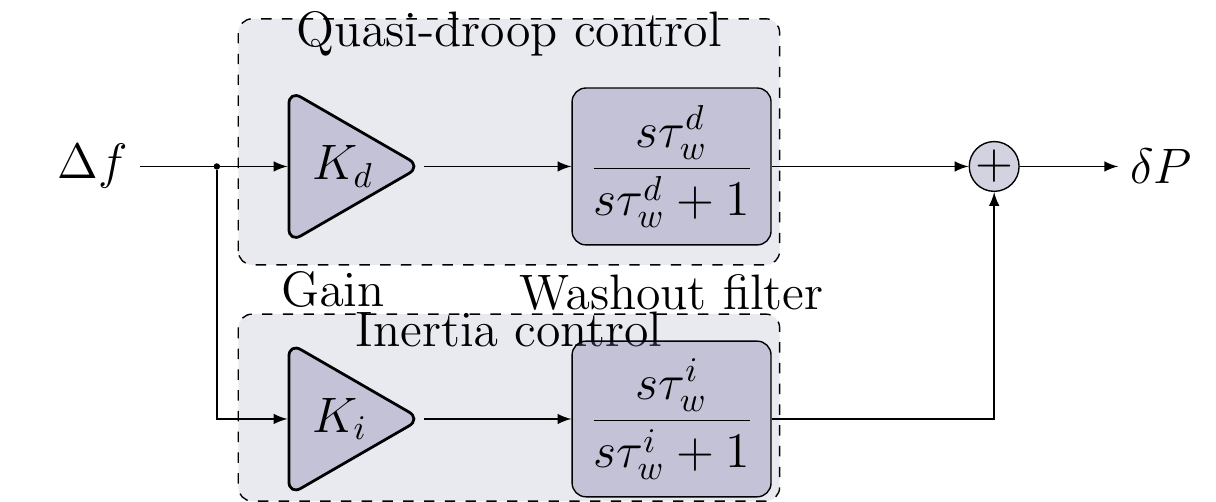}
\caption{Supercapacitor bank grid frequency control module}
\label{fig:scfreq}
\end{figure}

This block is shown in Fig. \ref{fig:scfreq}. The input to this block is the grid frequency signal estimated by the PLL and the output is the requested change in power. The type of implemented algorithm for frequency response can be arbitrary. However, based on the SC characteristics, in this study two control loops are used. The bottom loop is a standard virtual inertial response with a washout filter to make the output signal smoother since the time derivative operation inherently amplifies noise. The upper loop is more akin to a standard droop control, but it also has a washout filter which means this contribution will diminish in steady-state, hence the name \textit{quasi-droop}. 

The reasoning for this choice is the following: the SC does not have a lot of stored energy---if the standard droop control is employed then the SC output power is initially proportional to the frequency deviation. However, once the SC is discharged, the output power will fall to zero which will cause a bigger secondary frequency drop. By setting a large washout filter time constant, the output power will slowly diminish while the conventional units pick up. Therefore, the difference between the inertia control loop and quasi-droop control loop is in the washout filter time constant ($\tau_w^d \gg \tau_w^i$).

\section{Simulation and results}\label{sec:sim}
The performance of the proposed model is implemented and tested on a standard IEEE 14-bus test system included in DIgSILENT PowerFactory. Two scenarios are tested: loss of generating unit and low-voltage ride through event. Results obtained with the  ideal model and the proposed nonlinear model are compared for different capacitor sizes and initial conditions. 100 MW SC bank is connected to bus 06. Parameters of the SC bank system are given in the Appendix.

An actual supercapacitor system was not available for validation hence experimentally validated models were used for validation of model reduction. There is no complete control system model of SC in the open literature that can be easily reproduced in system studies. The model proposed here is modular and generic enough to be implemented in standard power system simulation software or modified if necessary. It can adequately capture the relevant SC dynamics as the modelled SC control is generic/flexible with the capacitor model based on past experimentally validated results \cite{Musolino2013, Faranda2007}.
\subsection{Loss of generating unit}
\begin{figure}[!t]
\centering
\subfloat[$C_{sc} = 900 + 40u_{C}$; $C_0/C_{max} = 0.9$]{\includegraphics[width=1.45in]{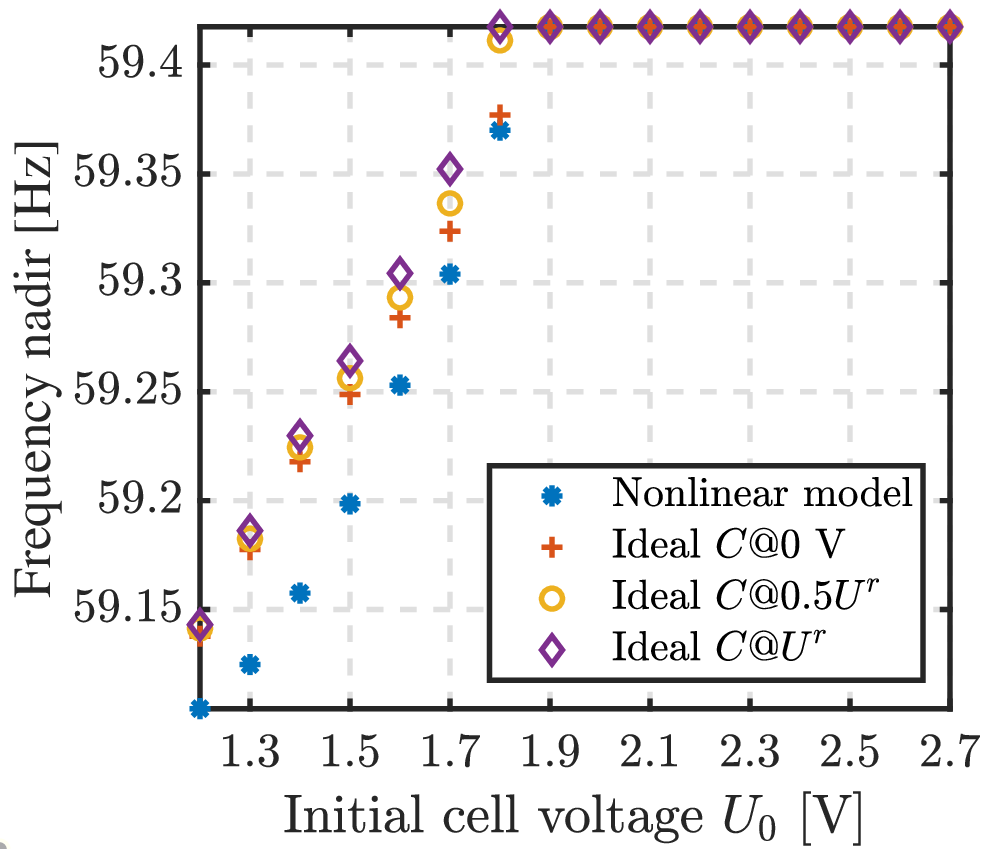}%
\label{fig:virnadir90}}
\hfil
\subfloat[$C_{sc} = 750 + 90u_{C}$; $C_0/C_{max} = 0.75$]{\includegraphics[width=1.45in]{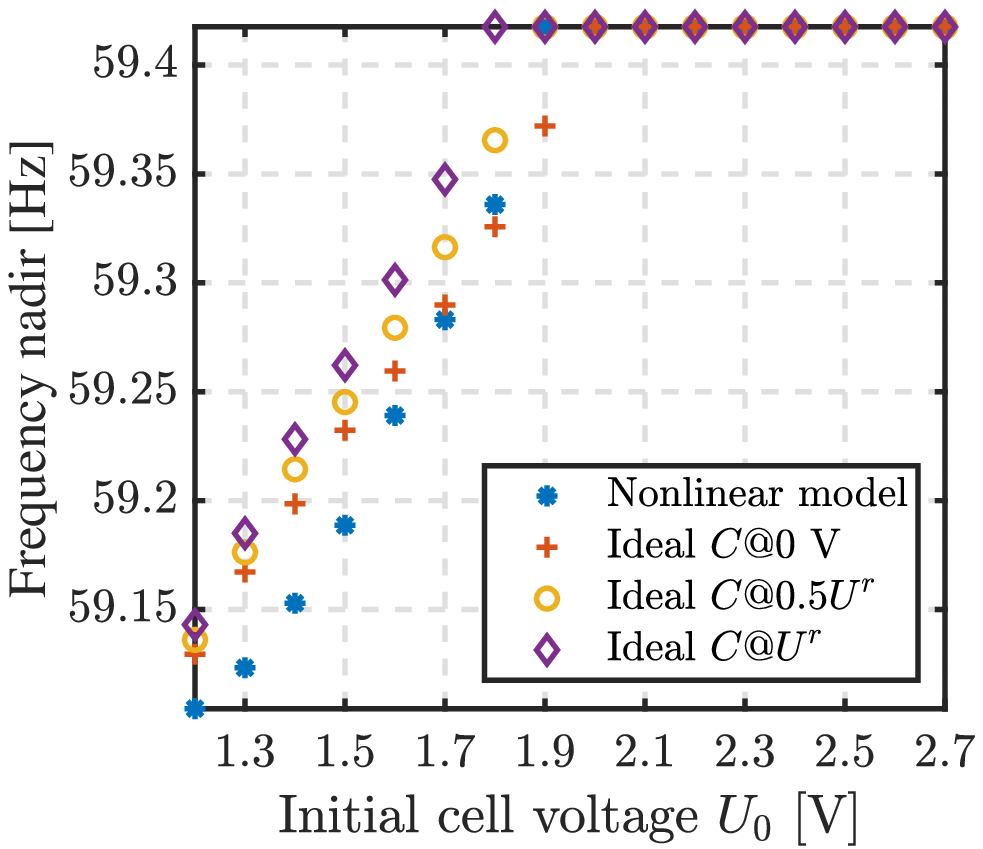}%
\label{fig:virnadir75}}
\hfil
\subfloat[$C_{sc} = 600 + 150u_{C}$; $C_0/C_{max} = 0.6$]{\includegraphics[width=1.45in]{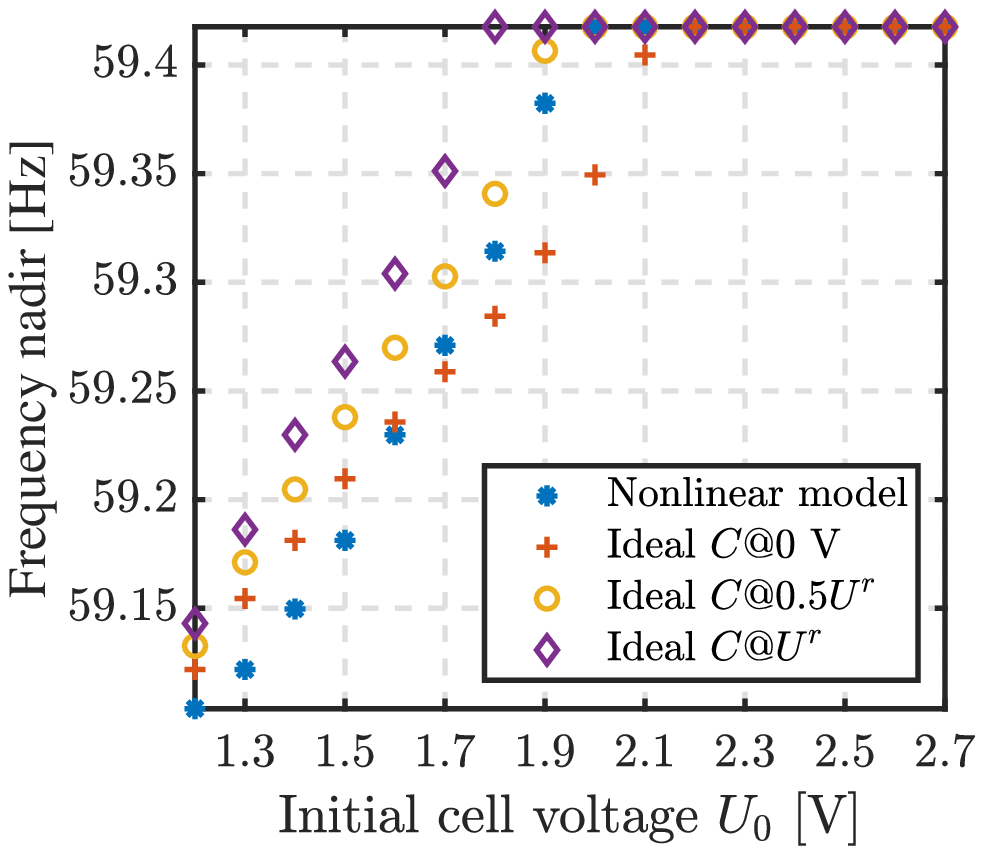}%
\label{fig:virnadir60}}
\hfil
\subfloat[Average RoCoF for $C_{sc} = 600 + 150u_{C}$]{\includegraphics[width=1.45in]{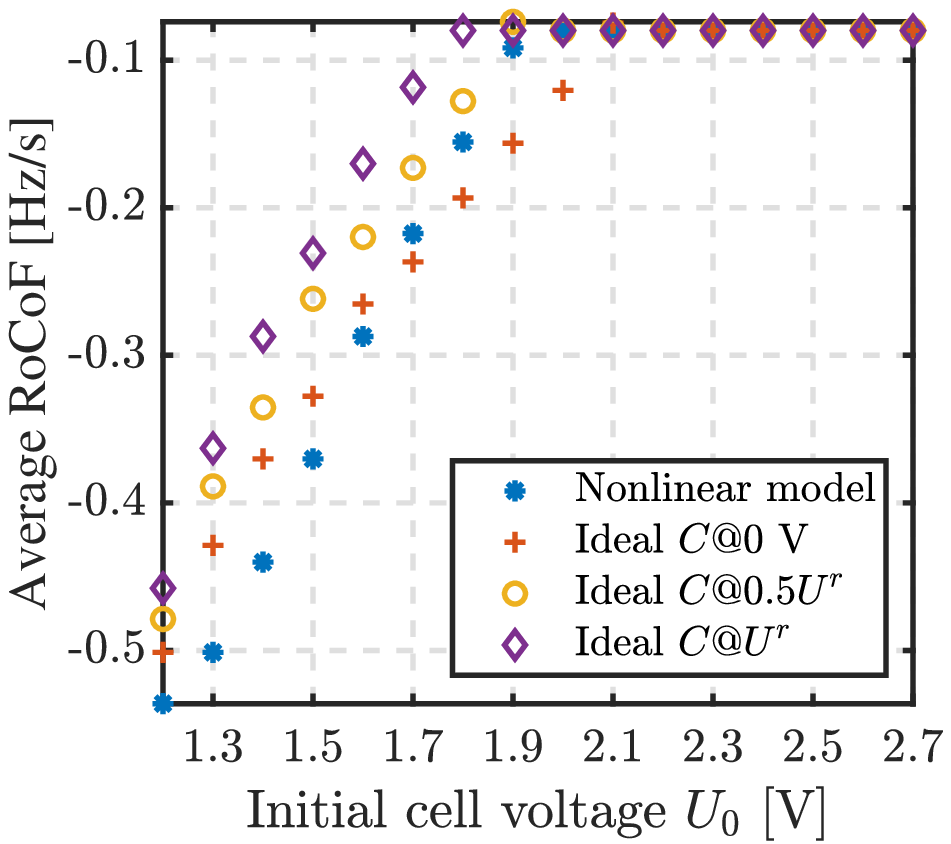}%
\label{fig:virrocof95MW600}}
\caption{Frequency nadir with different supercapacitor models for virtual inertial response control}
\label{fig:virnadir}
\end{figure}

\begin{figure}[!t]
\centering
\subfloat[$C_{sc} = 900 + 40u_{C}$; $C_0/C_{max} = 0.9$]{\includegraphics[width=1.45in]{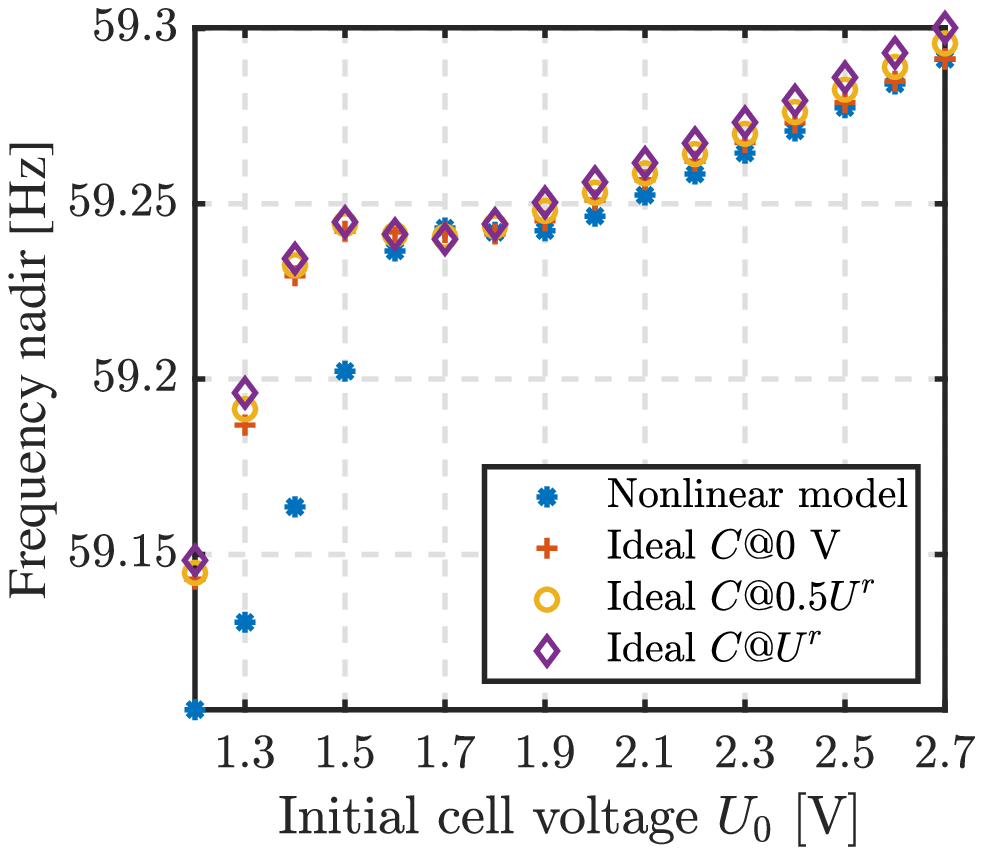}%
\label{fig:droopnadir90}}
\hfil
\subfloat[$C_{sc} = 600 + 150u_{C}$; $C_0/C_{max} = 0.6$]{\includegraphics[width=1.45in]{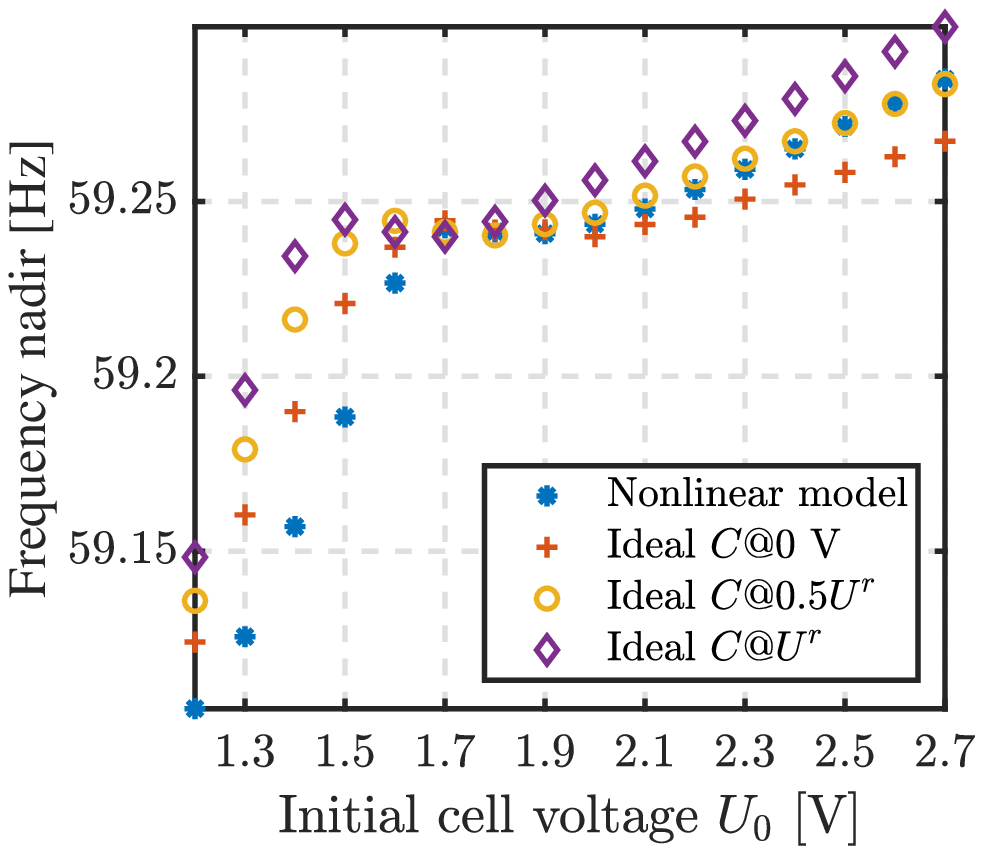}%
\label{fig:droopnadir60}}
\hfil
\subfloat[$C_{sc} = 900 + 40u_{C}$; $C_0/C_{max} = 0.9$]{\includegraphics[width=1.45in]{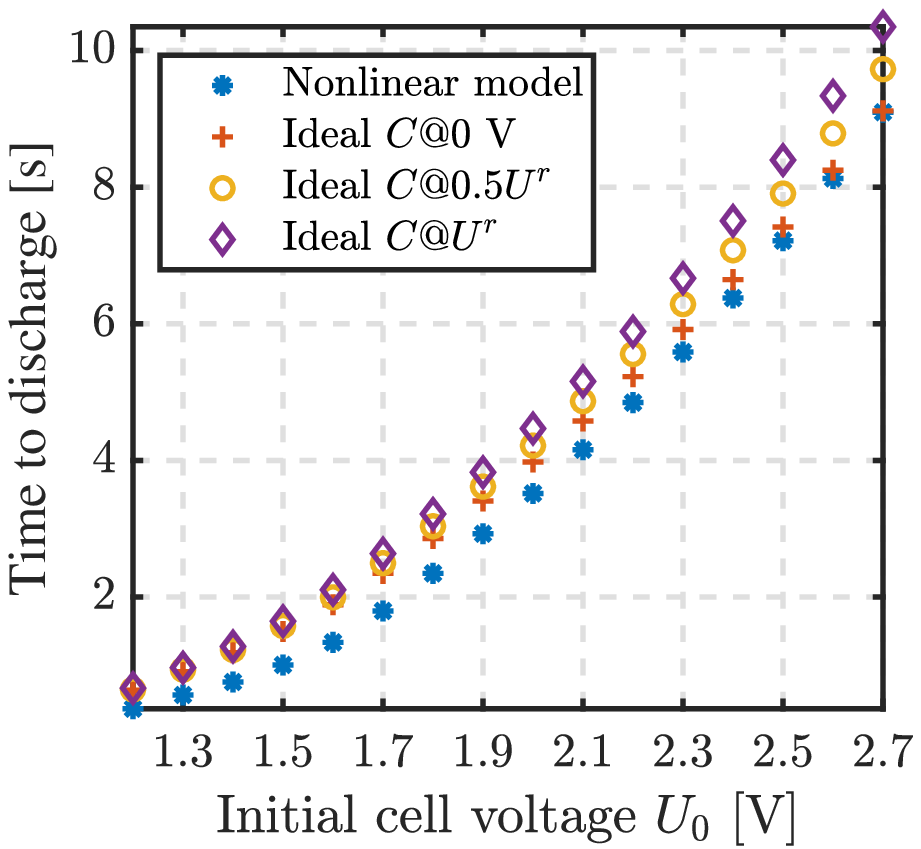}%
\label{fig:droopdt90}}
\hfil
\subfloat[$C_{sc} = 600 + 150u_{C}$; $C_0/C_{max} = 0.6$]{\includegraphics[width=1.45in]{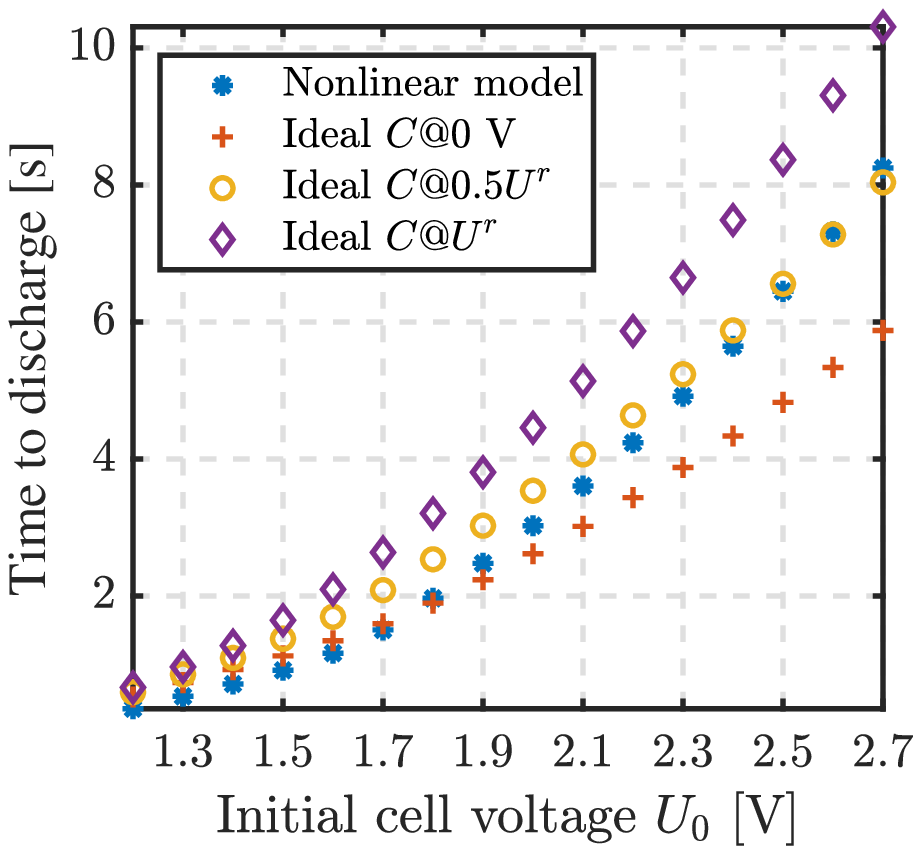}%
\label{fig:droopdt60}}
\caption{Frequency nadir and time to discharge with different supercapacitor models for quasi-droop control}
\label{fig:droopnadir}
\end{figure}

Two cases are analysed to trigger an underfrequency event: i) at $t = 1$ s, generator G02 delivering 95 MW is disconnected; ii) at $t = 1$ s, generator G02 delivering 190 MW is disconnected. The performance of the ideal model is compared to the performance of the nonlinear model. There are several parameters which may affect the performance of both models: initial capacitor voltage, ideal capacitor capacitance and how much the SC capacitance varies with voltage (i.e. ratio of capacitance at 0 V to capacitance at rated voltage). SC bank size is 100 MW. Characteristics of the system frequency response were observed and results for VIR control and quasi-droop control for 95 MW disturbance are shown separately in Fig. \ref{fig:virnadir} and Fig. \ref{fig:droopnadir}, respectively (shown separately because parameters of quasi-droop control impact the frequency response differently; nonetheless, the two control loops can be used simultaneously as shown in Fig. \ref{fig:scfreq}).

There is no significant difference in frequency nadir between using ideal and nonlinear model for a range of initial SC voltage values and capacitance models (Fig. \ref{fig:virnadir90}--\ref{fig:virnadir60}, \ref{fig:droopnadir90}--\ref{fig:droopnadir60}); however, the actual range depends on the size of the disturbance and control system parameters. Table \ref{tab:ranges} shows the operating range in terms of SoC in which the ideal model does not adequately represent the nonlinear model (inaccurate range) meaning that the range in which the ideal model is indeed adequate (accurate range) is complementary to the shown range. The criterion for selection of inaccurate range is the relative error between the nonlinear and ideal model for maximum frequency deviation (if this error is $>5$\%, then the ideal model does not adequately represent the nonlinear model). In Table \ref{tab:ranges}, best ideal model is highlighted by a shaded cell. Generally, for the same size of SC bank the ideal model is accurate  in a wider range for a smaller disturbance, and in a more narrow range for a larger distrubance. Furthermore, smaller gain and longer washout time constant of quasi-droop control increase the accurate range of the ideal model. In all cases, the inaccurate range is wider for larger capacitance of ideal model, although the values obtained with quasi-droop control are not as sensitive to the value of capacitance of the ideal model.

For a smaller disturbance, the inaccurate range is between 20\% and 50\% SoC for VIR control and between 20\% and 30\% for quasi-droop control meaning that the ideal model inaccurately represents the nonlinear model for low to medium SoC. When the disturbance exceeds the size of the SC, the inaccurate range is between 25\% and 70\% for VIR, although this range can be shifted to the higher SoC range if the variable part of SC capacitance is bigger (e.g. 71\% to 85\% SoC for 25\% variable capacitance and 70\% to 90\% SoC for 40\% variable capacitance). The inaccurate range for quasi-droop control is between 20\% and 40\% SoC. Note that for both disturbance sizes, there are empty cells in the table which mean that the ideal model in those cases accurately represents the nonlinear model for the whole operating range. One must keep in mind that the results are shown for an underfrequency event---for an overfrequency event the observed behaviour is complementary, i.e. the inaccurate range is in higher SoC because the capacitor is charging in that case and is limited by the maximum voltage.

Fig.\ref{fig:virrocof95MW600} shows average RoCoF for the SC with 40\% variable capacitance. It can be seen that the RoCoF increases with lower SoC because the SC bank will not be able to deliver the requested power before reaching minimum SoV. The RoCoF plots correspond to the frequency nadir plot (e.g. compare Fig. \ref{fig:virrocof95MW600} and Fig. \ref{fig:virnadir60}) for all SC models. Generally, larger variable capacitance will result in larger RoCoF and larger nadir for the same initial conditions (e.g. compare the nadir at 1.7 V in Fig. \ref{fig:virnadir}). The observed differences in nadir and RoCoF between different SC expressions are $<0.05$ Hz and $<0.05$ Hz/s, respectively.

Maximum possible difference in frequency nadir for all analysed scenarios is shown in Table \ref{tab:nadir}. In 34 out of 36 analysed cases, the ideal model with minimum capacitance will yield the best results (error is $\leq 0.1$ Hz) for both types of control. However, if the variable capacitance is larger (e.g. 40\% model), than the smallest maximum difference was achieved for the ideal model with average capacitance ($C@0.5U^r$).

Fig. \ref{fig:droopdt90} and Fig. \ref{fig:droopdt60} show the time to discharge of 10\% and 40\% variable capacitance models. The size of variable capacitance impacts which ideal model adequately describes the nonlinear model in terms of time to discharge. The mean absolute relative error in discharge time between the ideal model and the nonlinear model with 10\% variable capacitance is 16.4\% for ideal model with $C@U^r$, 12.8\% for ideal model with $C@0.5U^r$ and 8.7\% for ideal model with $C@0$ V. The same error for 40\% variable capacitance is 24.5\%, 9.9\% and 10.4\%, respectively. As the variable part of capacitance increases, so does the error of the ideal models with minimum and maximum capacitance and the average model will be the most accurate on average in terms of time to discharge taking into account the whole operating range.

Taking into account frequency nadir, RoCoF and time to discharge, the best ideal model for SC with up to 25\% variable capacitance is the ideal model with minimum capacitance ($C@0$ V) and this model is the same for both types of control. However, if the variable capacitance is larger than that, e.g. 40\%, the best ideal model may actually be the model with average capacitance for both types of control (for VIR control this is with respect to maximum absolute difference in frequency nadir while for quasi-droop control this is with respect to time to discharge).

\begin{table*}[t]
    \caption{Operating range (SoC) in which ideal model does not describe nonlinear model accurately with respect to maximum frequency deviation ($>\pm 5$\% error); narrower range is better}
    \label{tab:ranges}
    \centering
    \tiny
    \begin{tabular}{|c||c|c|c||c|c|c|}
        \hline
        \multicolumn{7}{|c|}{Designed operating range: 1.1 V -- 2.7 V (40\%--100\% $U_r$ $\sim$ 15\%--100\% SoC)}\\
        \hline
        \multicolumn{7}{|c|}{Disturbance size: 95 MW}\\
        \hline
         \multirow{2}{*}{Model} & \multicolumn{3}{c||}{VIR control} & \multicolumn{3}{c|}{Quasi-droop control} \\
         \cline{2-7}
         & $C@0$ V & $C@0.5U_r$ & $C@U_r$ & $C@0$ V & $C@0.5U_r$ & $C@U_r$\\
         \hline
         $C_0/C_{max} = 0.9$ & \cellcolor[gray]{.8}{22\%--30\%} & 22\%--43\% & 22\%--43\% & \cellcolor[gray]{.8}{22\%--30\%} & 22\%--30\% & 22\%--30\% \\
         \hline
         $C_0/C_{max} = 0.75$ &\cellcolor[gray]{.8}{21\%--28\%} & 21\%--33\% & 21\%--42\% & \cellcolor[gray]{.8}{21\%--28\%} & 21\%--28\% & 21\%--28\% \\
         \hline
         $C_0/C_{max} = 0.6$ & \cellcolor[gray]{.8}{40\%--50\%} & 20\%--35\% & 20\%--45\% & \cellcolor[gray]{.8}{-} & 20\%--27\%& 16\%--27\% \\
         \hline
        \multicolumn{7}{|c|}{Disturbance size: 190 MW}\\
        \hline
         \multirow{2}{*}{Model} & \multicolumn{3}{c||}{VIR control} & \multicolumn{3}{c|}{Quasi-droop control} \\
         \cline{2-7}
         & $C@0$ V & $C@0.5U_r$ & $C@U_r$ & $C@0$ V & $C@0.5U_r$ & $C@U_r$\\
         \hline
         $C_0/C_{max} = 0.9$ & \cellcolor[gray]{.8}{26\%--43\%} & 26\%--48\% & 26\%--66\% & \cellcolor[gray]{.8}{26\%--39\%} & 21\%--39\% & 21\%--39\% \\
         \hline
         $C_0/C_{max} = 0.75$ &\cellcolor[gray]{.8}{71\%--85\%} & 25\%--47\% & 25\%--71\% & \cellcolor[gray]{.8}{-} & 25\%--42\% & 21\%--42\% \\
         \hline
         $C_0/C_{max} = 0.6$ & \cellcolor[gray]{.8}{70\%--90\%} & 23\%--45\% & 23\%--70\% & \cellcolor[gray]{.8}{-} & 23\%--40\%& 20\%--40\% \\
         \hline
    \end{tabular}
\end{table*}

\begin{table*}[t]
    \caption{Maximum absolute difference in frequency nadir for all analysed cases (smaller is better)}
    \label{tab:nadir}
    \centering
    \tiny
    \begin{tabular}{|c||c|c|c||c|c|c|}
    \hline
        \multicolumn{7}{|c|}{Disturbance size: 95 MW}\\
        \hline
         \multirow{2}{*}{Model} & \multicolumn{3}{c||}{VIR control} & \multicolumn{3}{c|}{Quasi-droop control} \\
         \cline{2-7}
         & $C@0$ V & $C@0.5U_r$ & $C@U_r$ & $C@0$ V & $C@0.5U_r$ & $C@U_r$\\
         \hline
         $C_0/C_{max} = 0.9$ & \cellcolor[gray]{.8}{0.060 Hz} & 0.067 Hz & 0.072 Hz & \cellcolor[gray]{.8}{0.066 Hz} & 0.069 Hz & 0.071 Hz \\
         \hline
         $C_0/C_{max} = 0.75$ &\cellcolor[gray]{.8}{0.046 Hz} & 0.062 Hz & 0.082 Hz & \cellcolor[gray]{.8}{0.050 Hz} & 0.065 Hz & 0.074 Hz \\
         \hline
         $C_0/C_{max} = 0.6$ & 0.069 Hz & \cellcolor[gray]{.8}{0.057} Hz & 0.103 Hz & \cellcolor[gray]{.8}{0.035 Hz} & 0.059 Hz & 0.077 Hz\\
         \hline
        \multicolumn{7}{|c|}{Disturbance size: 190 MW}\\
        \hline
         \multirow{2}{*}{Model} & \multicolumn{3}{c||}{VIR control} & \multicolumn{3}{c|}{Quasi-droop control} \\
         \cline{2-7}
         & $C@0$ V & $C@0.5U_r$ & $C@U_r$ & $C@0$ V & $C@0.5U_r$ & $C@U_r$\\
         \hline
         $C_0/C_{max} = 0.9$ & \cellcolor[gray]{.8}{0.096 Hz} & 0.104 Hz & 0.117 Hz & \cellcolor[gray]{.8}{0.101 Hz} & 0.113 Hz & 0.124 Hz \\
         \hline
         $C_0/C_{max} = 0.75$ &\cellcolor[gray]{.8}{0.089 Hz} & 0.101 Hz & 0.134 Hz & \cellcolor[gray]{.8}{0.080 Hz} & 0.102 Hz & 0.130 Hz \\
         \hline
         $C_0/C_{max} = 0.6$ & 0.149 Hz & \cellcolor[gray]{.8}{0.094} Hz & 0.167 Hz & \cellcolor[gray]{.8}{0.065 Hz} & 0.099 Hz & 0.137 Hz \\
         \hline
    \end{tabular}
\end{table*}

The observed differences are important to accurately predict the performance of a SC bank during all operating points and therefore the grid frequency dynamics during large load-generation disturbances. For example, if the frequency is to be contained after the loss of a generator, using an ideal model might lead to a conclusion that the frequency will indeed be contained, while in reality, the SC will not be able to deliver necessary power output and underfrequency load-shedding (UFLS) will be triggered (e.g., see Fig. \ref{fig:virnadir90} and Fig. \ref{fig:droopnadir90} at 1.5 V if the first stage of UFLS is set at 59.2 Hz). SC can sustain an output power profile for a limited amount of time and it is necessary to accurately estimate the stored energy in order to guarantee the requested provision of system services. Choosing an inadequate capacitance value may lead to a difference in calculated frequency of over 0.1 Hz (e.g. Fig. \ref{fig:virnadir60} capacitance at rated voltage) as well the RoCoF (Fig. \ref{fig:virrocof95MW600}).

\subsection{Discussion on power and energy of a supercapacitor for a loss of generating unit}

An ideal capacitor will extract the maximum amount of energy between two voltage levels without any losses, however in a real supercapacitor a part of the stored energy will be dissipated as a thermal loss that is time-variant as the discharging current changes to maintain the requested power profile. Thus, the effective capacitance of an ideal model would have to be recalculated for each operating point which is not practical and since the real supercapacitor is nonlinear and the requested power profile depends on the external factors (i.e. size of disturbance, grid dynamics and frequency controller tuning) this is also practically impossible to do both analytically and apriori. In a general case, the ideal model will behave identically to the realistic model in terms of output power profile if the energy requested during the transient is less than the minimum of stored effective energy in an ideal model and in realistic model:
\begin{equation}
    E_{\text{tran}} < \min\{E_{\text{ideal}}, E_{\text{real}}\}
\end{equation}

Considering the same loss of a 95 MW generating unit G02 at $t=1$ s, the power profile and SoC are observed for a fully (100\% SoC) and partially ($\sim 45\%$ SoC) charged supercapacitor for different ideal model representations. Frequency controller parameters are set to $K_i = 100$ p.u., $K_d = 50$ p.u., $\tau_w^i = 1$ s, $\tau_w^d = 30$ s to induce a complete discharge of the supercapacitor bank.

\begin{figure}[!t]
\centering
\subfloat[Power profile for full charge]{\includegraphics[width=1.45in]{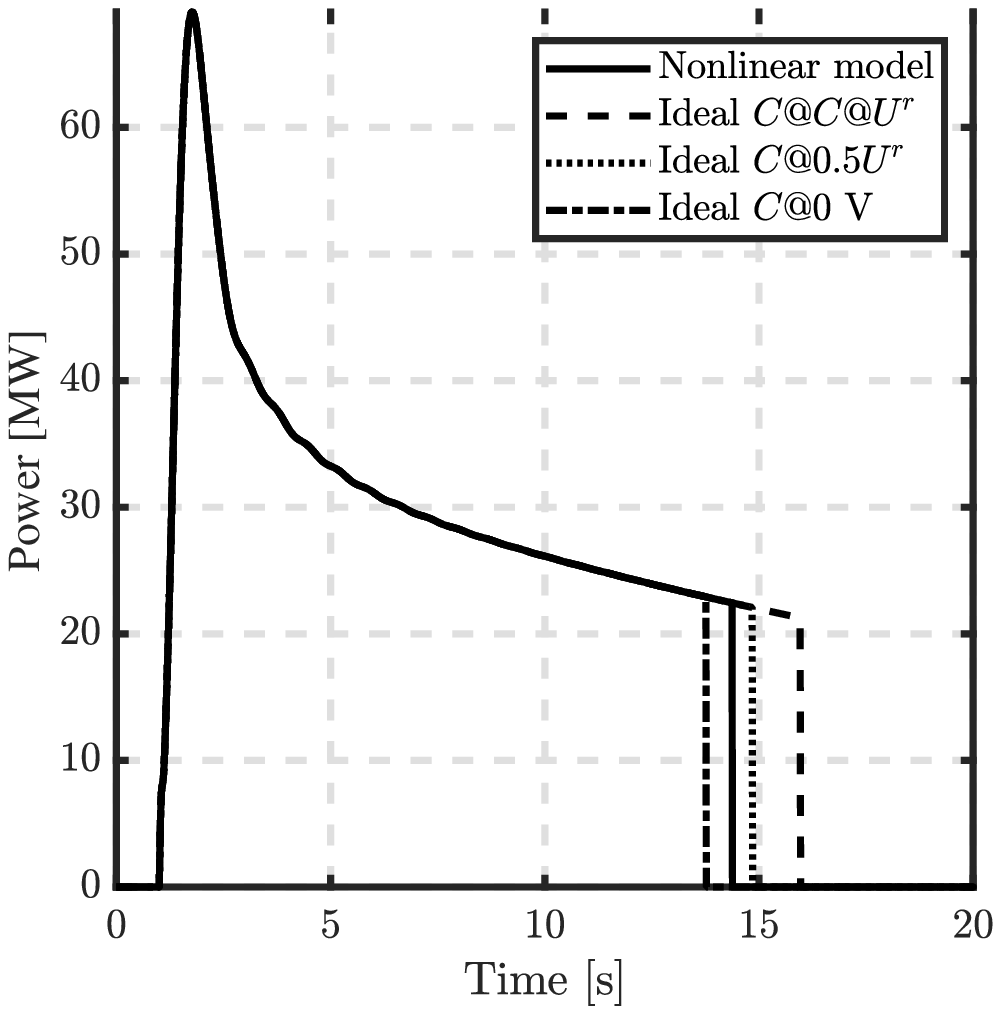}%
\label{fig:power_900_27}}
\hfil
\subfloat[SoC profile for full charge]{\includegraphics[width=1.45in]{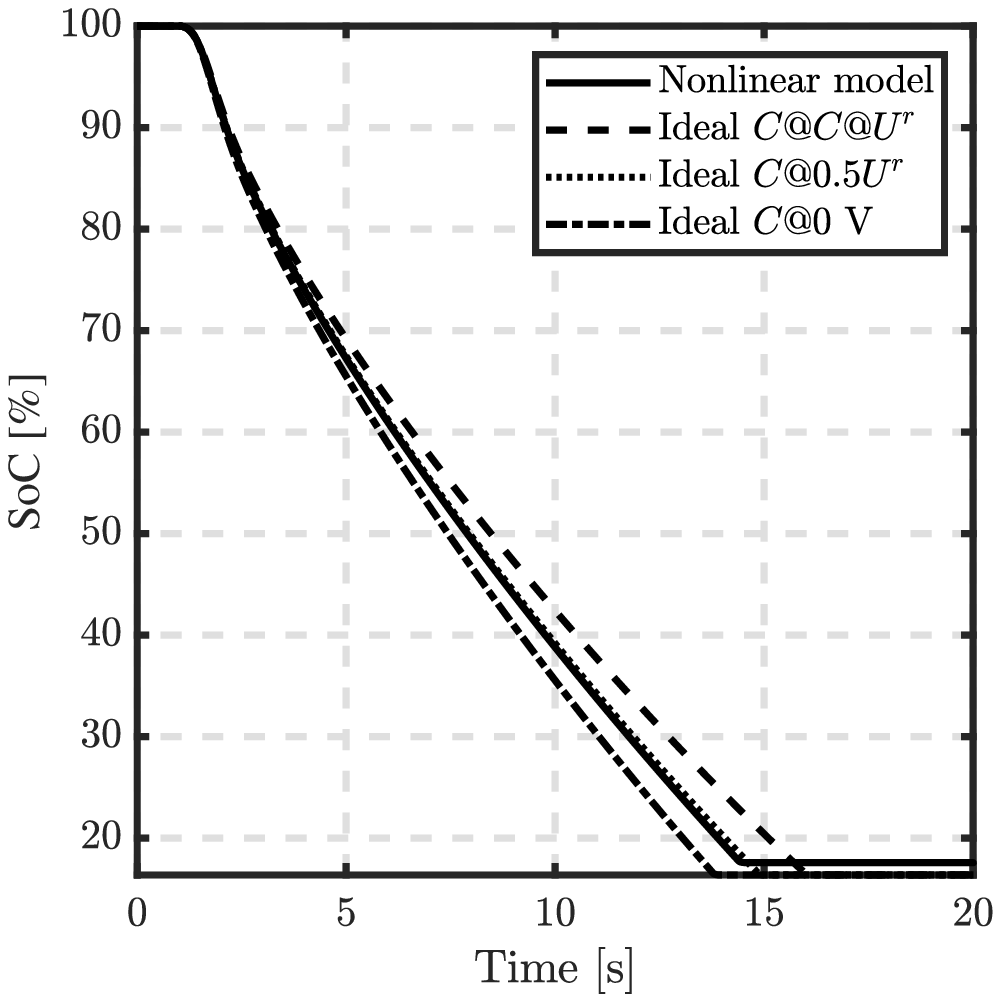}%
\label{fig:soc_900_27}}
\hfil
\subfloat[Power profile for partial charge]{\includegraphics[width=1.45in]{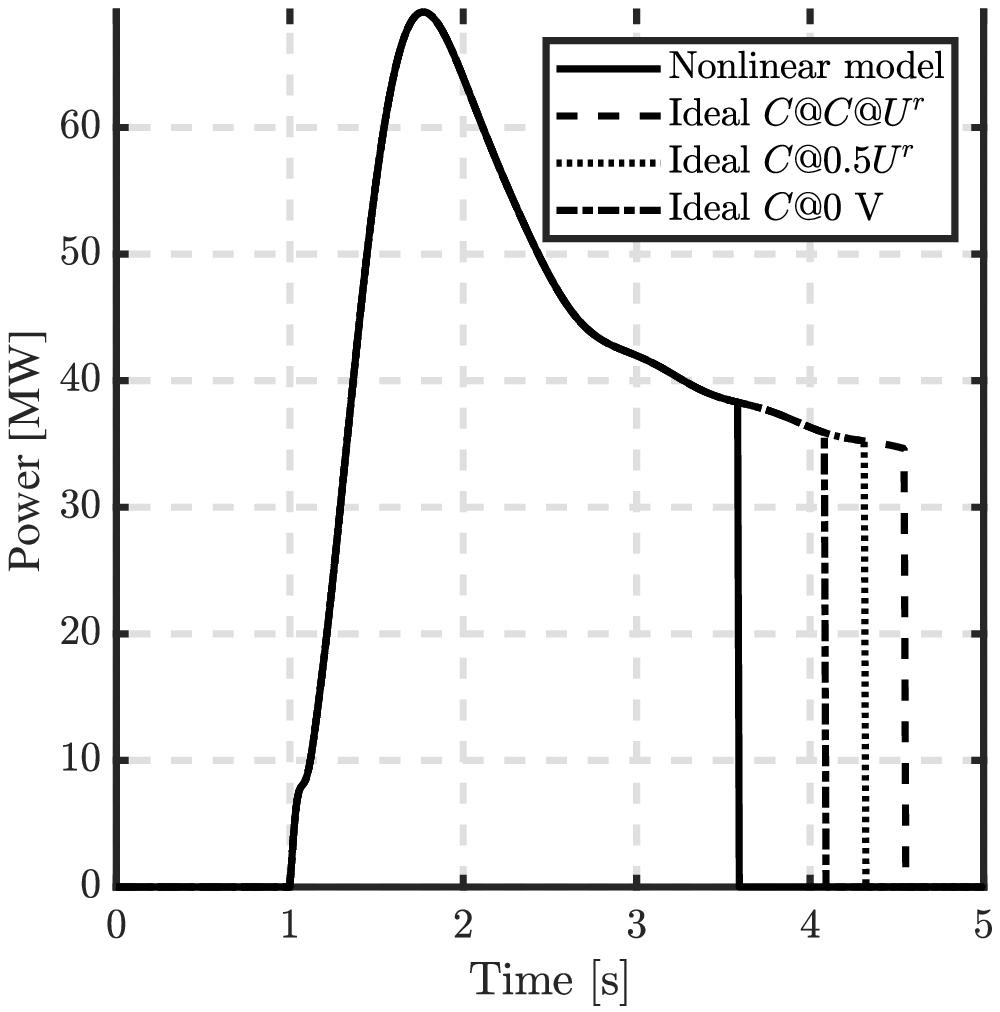}%
\label{fig:power_900_18}}
\hfil
\subfloat[SoC profile for partial charge]{\includegraphics[width=1.45in]{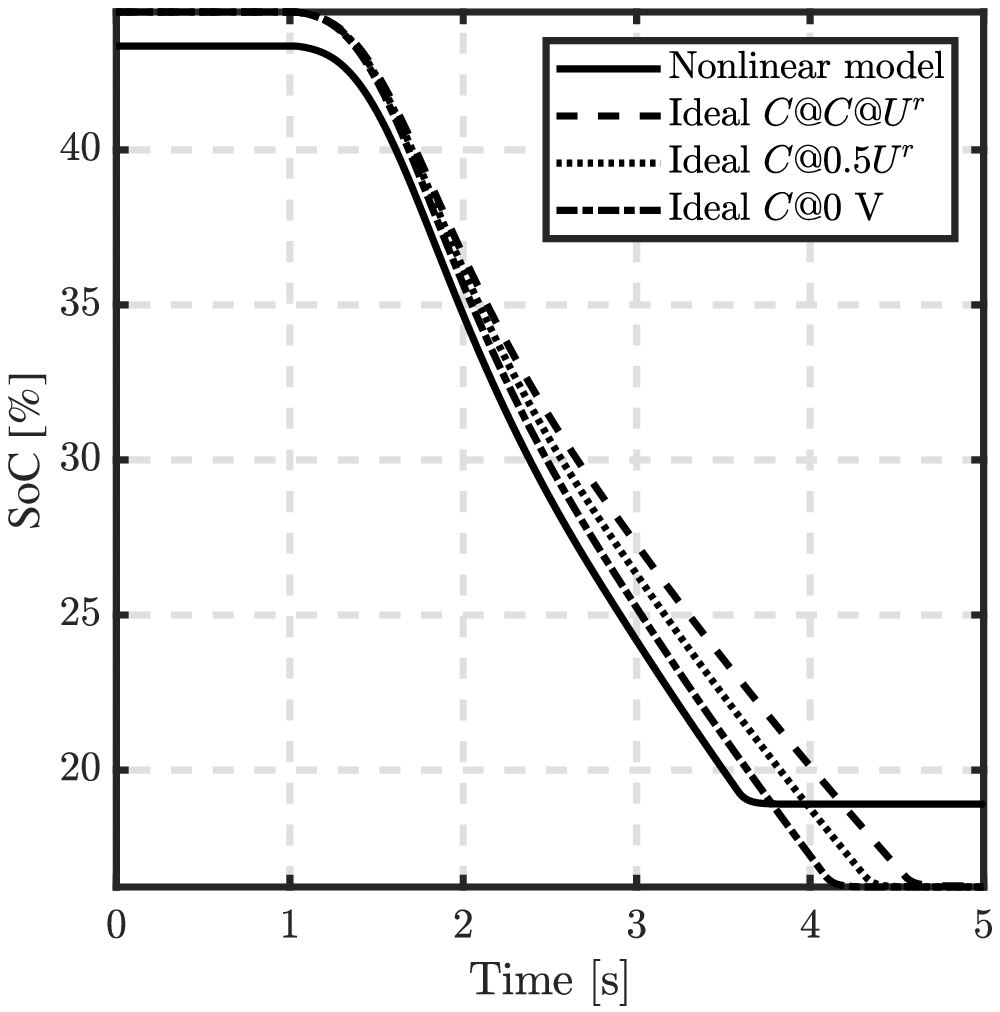}%
\label{fig:soc_900_18}}
\caption{Power and SoC profiles for fully charged (top) and partially charged (bottom) supercapacitor for 10\% variable capacitance ($C_{sc} = 900 + 40u_{C}$)}
\label{fig:powersoc900}
\end{figure}

Fig. \ref{fig:powersoc900} shows the results for a supercapacitor with 10\% variable capacitance. It can be seen that in the case of a fully charged supercapacitor, the model with average capacitance is the most accurate in terms of power profile and SoC (Fig. \ref{fig:power_900_27} and Fig. \ref{fig:soc_900_27}): ideal model with average capacitance can sustain the output power only for a fraction of a second longer than nonlinear model (14.85 s compared to 14.38, or 3.3\%). On the other hand, ideal model with rated capacitance sustains the output power 1.6 s longer (15.99s compared to 14.38 s or 11.2\% longer), while the ideal model with minimum capacitance sustains the output power 0.6 s shorter (13.77 s compared to 14.38 s, or -4.2\%). Fig. \ref{fig:soc_900_27} shows that energy depletes faster in the case of ideal model with minimum capacitance and slower in the case of ideal model with maximum capacitance, while the ideal model with average capacitance is fairly accurate in this case. When the supercapacitor is only partially charged when the disturbance happens, the differences are more pronounced (Fig. \ref{fig:power_900_18} and Fig. \ref{fig:soc_900_18}) and all ideal models overestimate the available energy: ideal models with minimum, average and rated capacitance sustain the output power for 0.5 s (+13.9\%), 0.74 s (+20.6\%) and 0.98 s (+27.3\%) longer, respectively. Furthermore, initial SoC is overestimated by ideal models by 1.1\%, it depletes slower and total change of SoC is 15.4\% bigger using ideal models.

\begin{figure}[!t]
\centering
\subfloat[Power profile for full charge]{\includegraphics[width=1.45in]{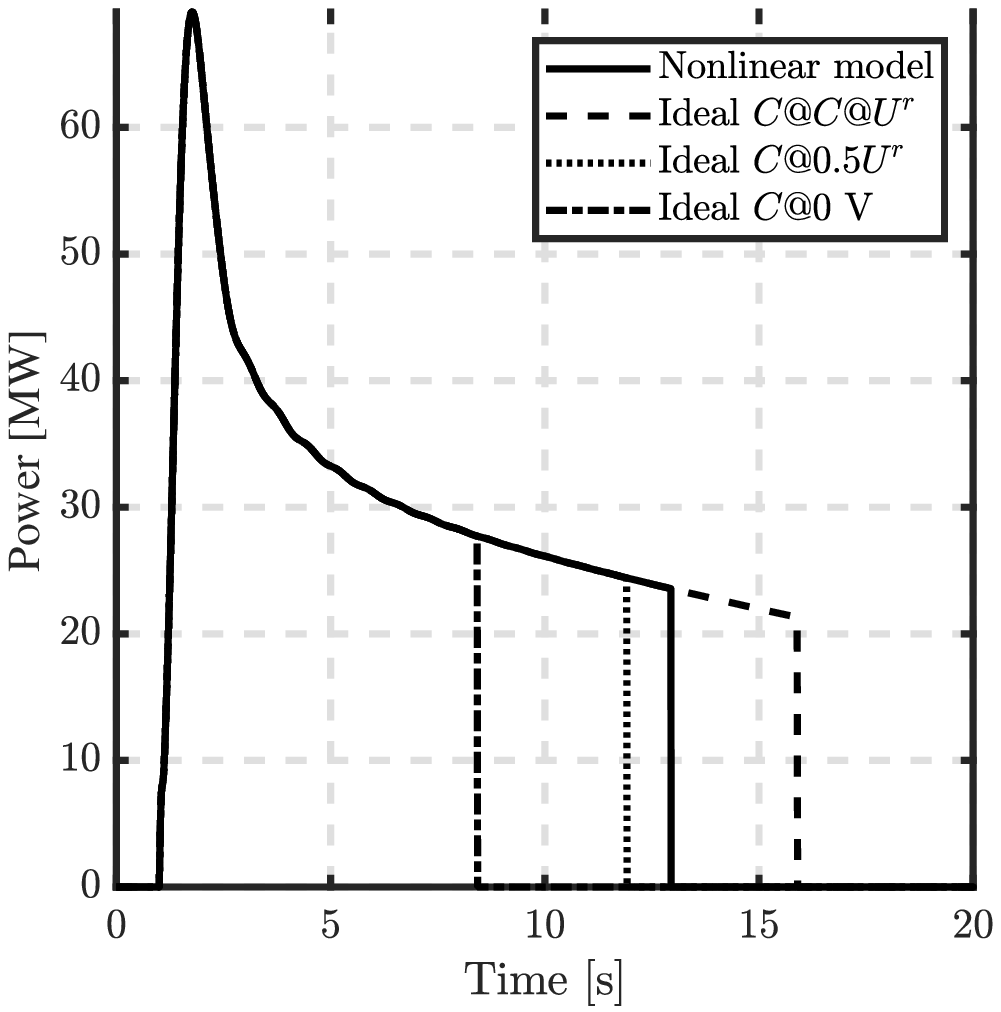}%
\label{fig:power_600_27}}
\hfil
\subfloat[SoC profile for full charge]{\includegraphics[width=1.45in]{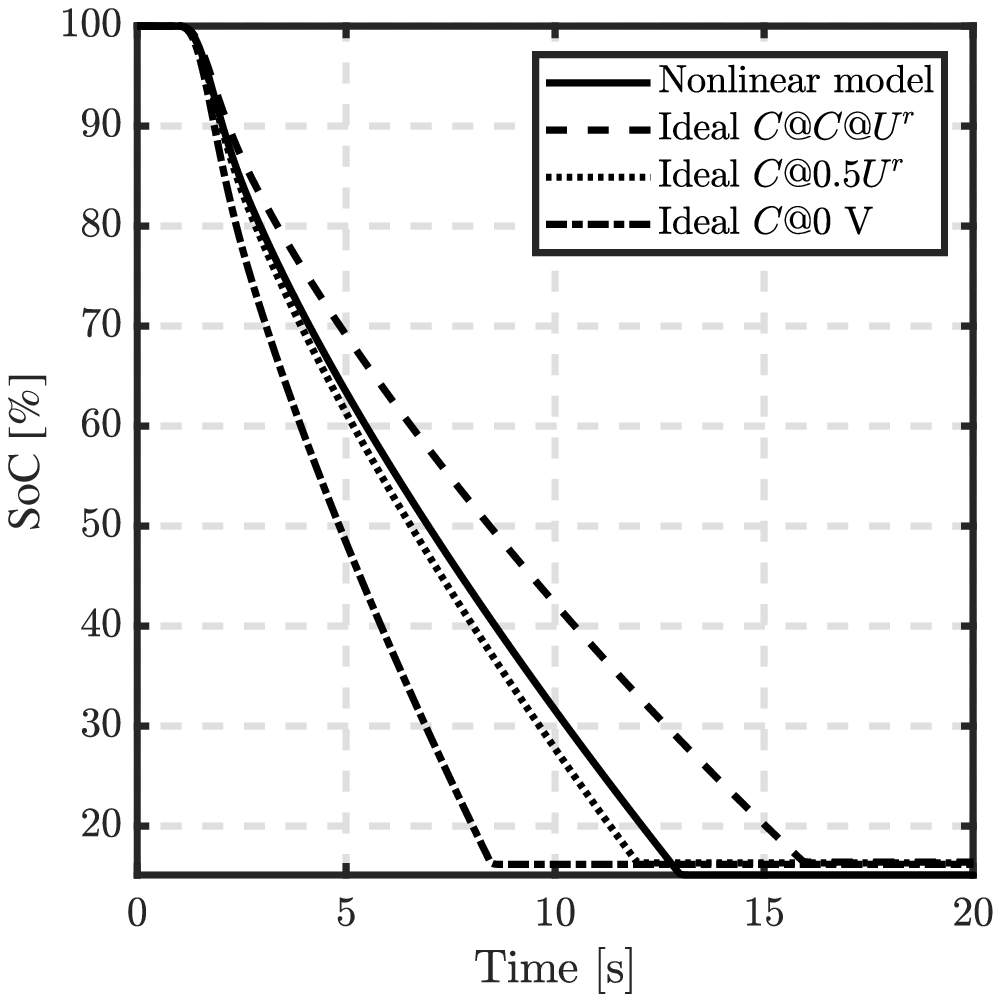}%
\label{fig:soc_600_27}}
\hfil
\subfloat[Power profile for partial charge]{\includegraphics[width=1.45in]{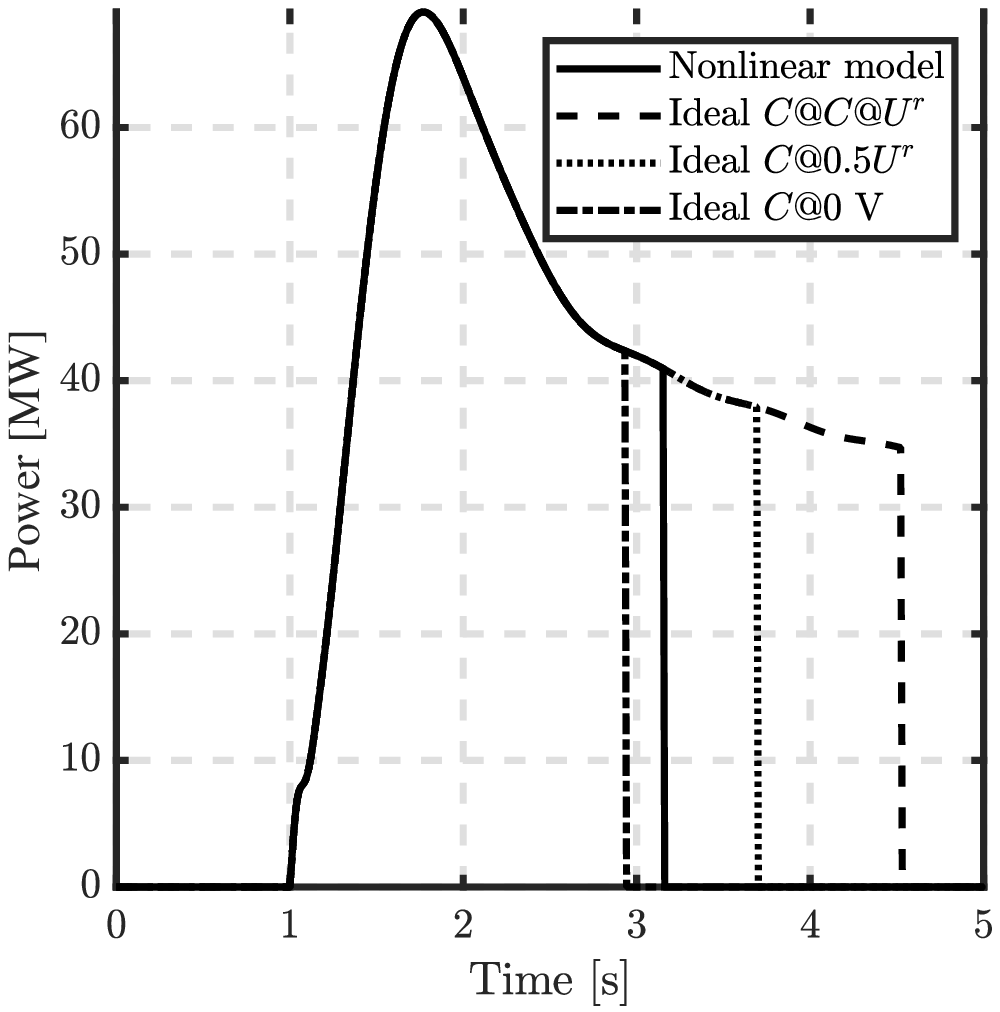}%
\label{fig:power_600_18}}
\hfil
\subfloat[SoC profile for partial charge]{\includegraphics[width=1.45in]{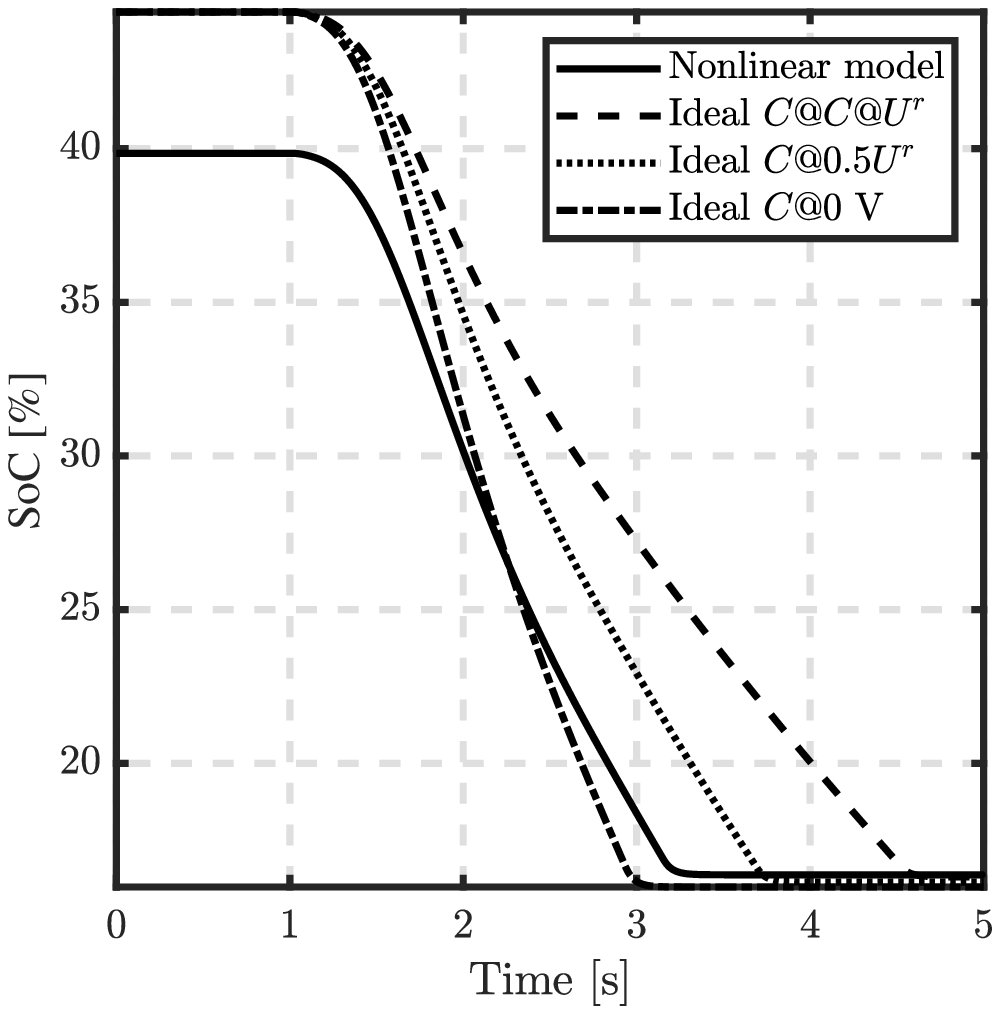}%
\label{fig:soc_600_18}}
\caption{Power and SoC profiles for fully charged (top) and partially charged (bottom) supercapacitor for 40\% variable capacitance ($C_{sc} = 600 + 150u_{C}$)}
\label{fig:powersoc600}
\end{figure}

Fig. \ref{fig:powersoc600} shows the results for a supercapacitor with 40\% variable capacitance. In the case of a fully charged the most accurate ideal model in terms of power profile and SoC is again the ideal model with average capacitance (Fig. \ref{fig:power_600_27} and Fig. \ref{fig:soc_600_27}): it can sustain the output power profile for 1 s shorter (11.88 s compared to 12.95 s or -8.3\%), while the ideal model with minimum capacitance can sustain the output power for 4.5 s shorter (8.43 s compared to 12.95 s, or -35.1\%) and the ideal model with maximum capacitance can sustain the output power for 3 s longer (15.9 s compared to 12.95 s, or +22.8\%). Fig. \ref{fig:soc_600_27} shows that the SoC can deplete at significantly different rates depending on the choice of ideal capacitance. For a partially charged supercapacitor (Fig. \ref{fig:power_600_18} and Fig. \ref{fig:soc_600_18}), the most accurate ideal model with respect to power profile and SoC is the one with minimum capacitance: it discharges 0.22 s earlier (-7.0\%) than the nonlinear model. Ideal model with average and rated capacitance overestimate the stored energy and they can sustain the power output for a longer time: 0.54 s longer (+17.1\%) and 1.37 s longer (+43.3\%). Ideal model overestimates the initial SoC by 4.6\% and the total change of SoC is 21.4\% bigger using ideal model.

It can be concluded that as the variable part of capacitance increases, there is a bigger error in estimation of initial SoC as well as in depletion rate of energy between ideal models and nonlinear model. Also, as the initial SoC decreases, different ideal models describe the nonlinear model better. Therefore, the nonlinear voltage dynamics and losses of a real supercapacitor cell make the modelling using ideal or simplified models problematic.

\subsection{Low voltage ride through}
\begin{figure}[!t]
\centering
\subfloat[$C_{sc} = 900 + 40u_{C}$; $C_{ideal} = 900$ F]{\includegraphics[width=1.5in]{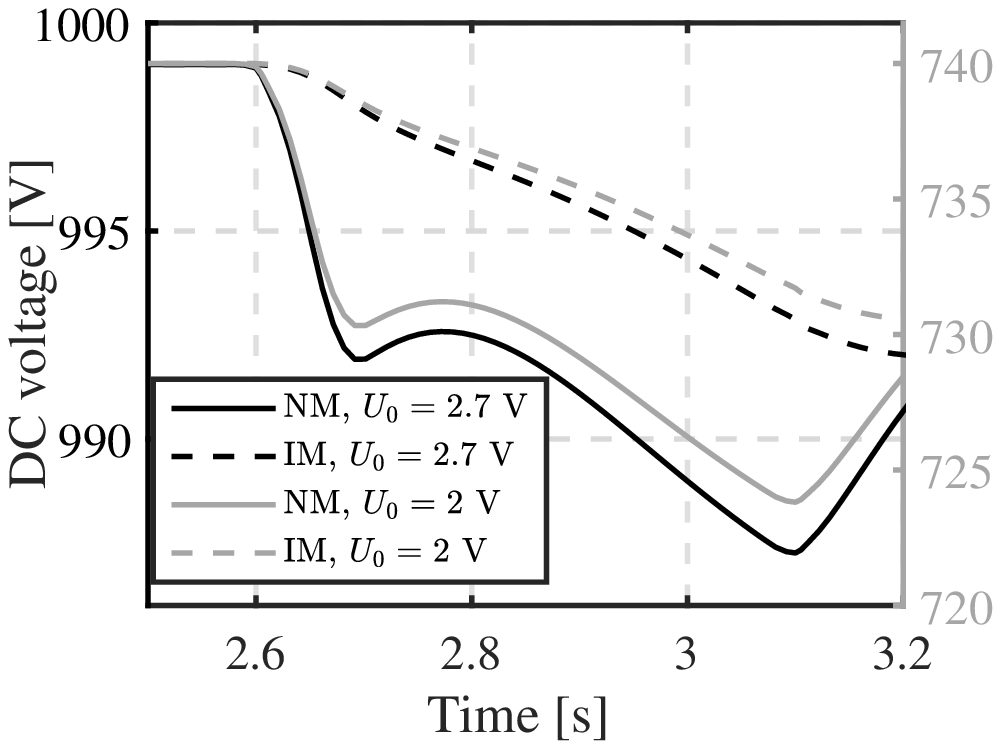}%
\label{fig:lvrtdc900}}
\hfil
\subfloat[$C_{sc} = 600 + 150u_{C}$; $C_{ideal} = 600$ F]{\includegraphics[width=1.5in]{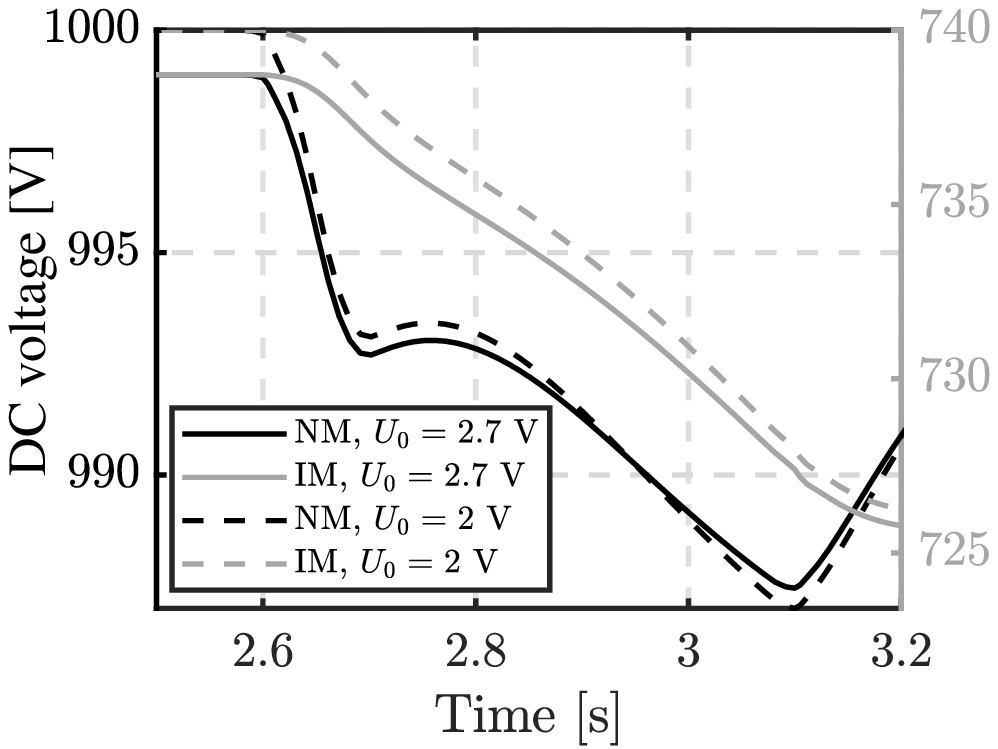}%
\label{fig:lvrtdc600}}
\caption{DC voltage of a supercapacitor bank during fault for different initial voltages and supercapacitor models}
\label{fig:lvrtdc}
\end{figure}

Performance of different models has also been evaluated for balanced three-phase faults near Bus 06. In the first scenario, a 500 ms three-phase self-clearing fault has been applied to Bus 02 and Fig. \ref{fig:lvrtdc} shows the SC DC voltage for different initial voltages and SC models. In all cases, the initial voltage and SC voltage dependence do not influence the model performance significantly. The difference between the ideal and nonlinear model (IM,NM) is more visible at the first glance: the nonlinear model has a greater voltage change due to the voltage drop on the ESR. However, maximum difference is less than 10 V so it can also be neglected.

\begin{figure}[!t]
\centering
\subfloat[$C_{sc} = 900 + 40u_{C}$; $C_{ideal} = 900$ F]{\includegraphics[width=1.5in]{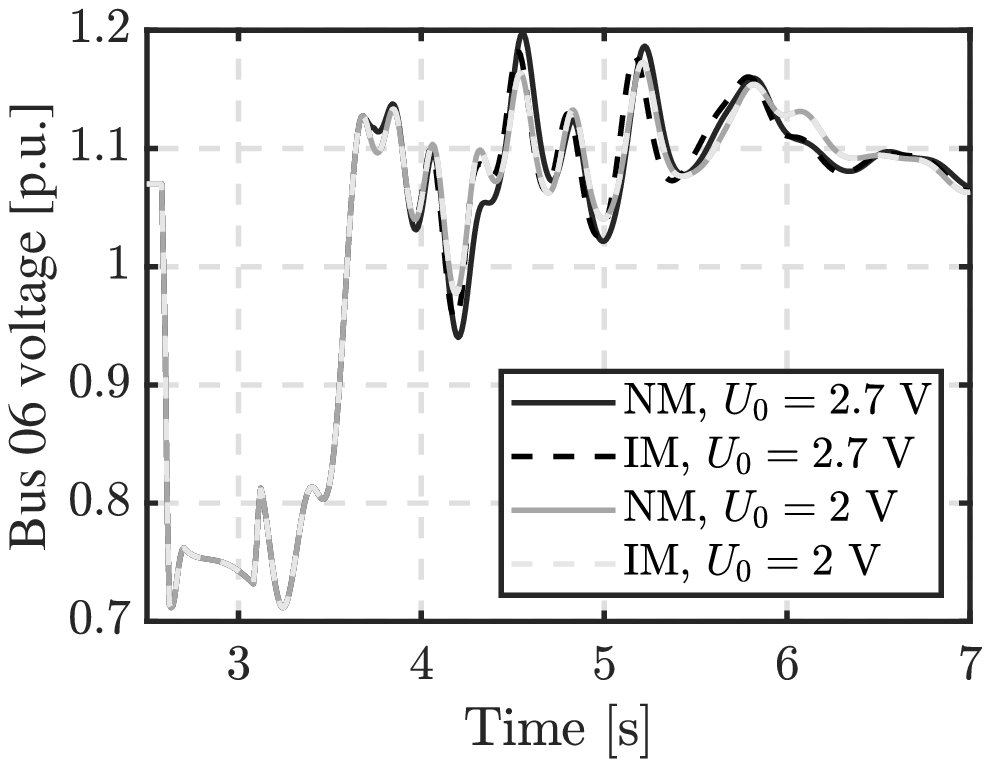}%
\label{fig:lvrtac900}}
\hfil
\subfloat[$C_{sc} = 600 + 150u_{C}$; $C_{ideal} = 600$ F]{\includegraphics[width=1.5in]{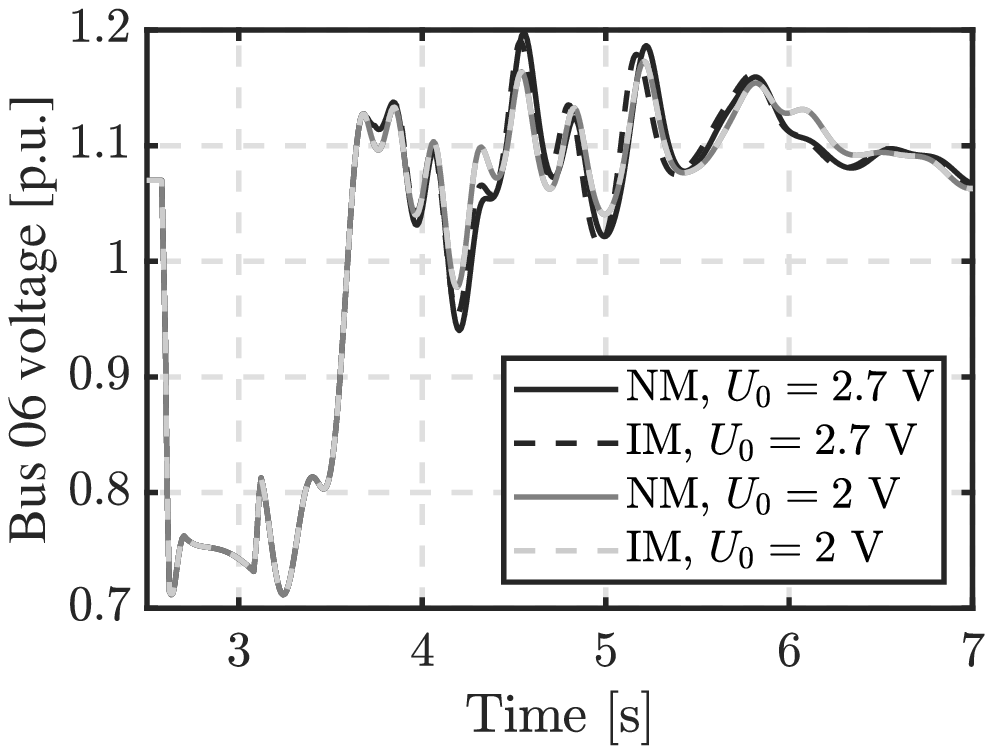}%
\label{fig:lvrtac600}}
\caption{AC voltage at Bus 06 after a disturbance}
\label{fig:lvrtac}
\end{figure}

Bus 06 AC voltage profile is shown in Fig. \ref{fig:lvrtac}. Similarly, the initial voltage, SC voltage dependence and the type of model do not have a significant impact on the post-fault voltage transient, while the profiles are identical during the fault. Maximum observed difference of absolute voltage values is 0.03 p.u. between nonlinear and ideal model (both fully charged) for both SC capacitance expressions. For a partially charged SC ($U_0 = 2$ V), this difference is 0 for both capacitance expressions. This is because the system is initially accelerating after the fault clears, so the SC is charging and the fully charged SC is the worst case scenario since it can't accept much more charge (a small tolerance of $<1\%$ exists between the rated voltage and cut-off voltage). Therefore, SC with low SoC will behave identically to the partially charged SC (e.g., $U_0 = 2$ V or about 50\% SoC). Maximum observed difference of absolute voltage values between partially charged and fully charged SC is 0.07 p.u.

\begin{figure}[!t]
\centering
\subfloat[$C_{sc} = 600 + 150u_{C}$; $C_{ideal} = 600$ F]{\includegraphics[width=1.5in]{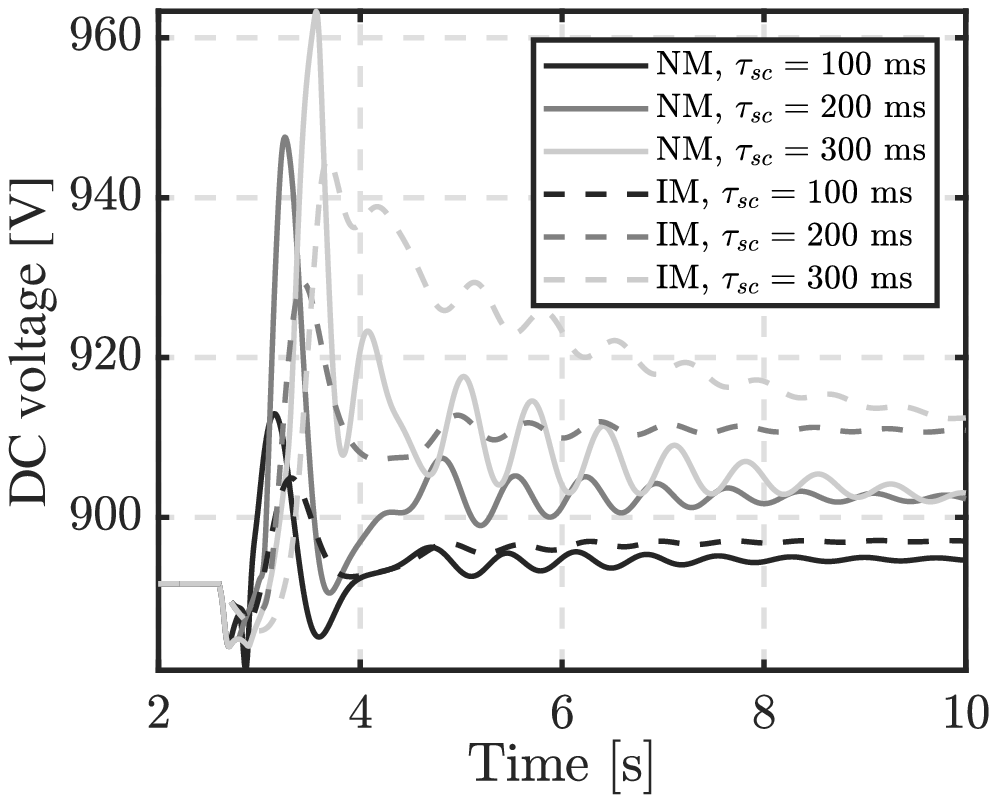}%
\label{fig:lvrt-sctime}}
\hfil
\subfloat[$C_{sc} = 600 + 150u_{C}$; $C_{ideal} = 600$ F]{\includegraphics[width=1.5in]{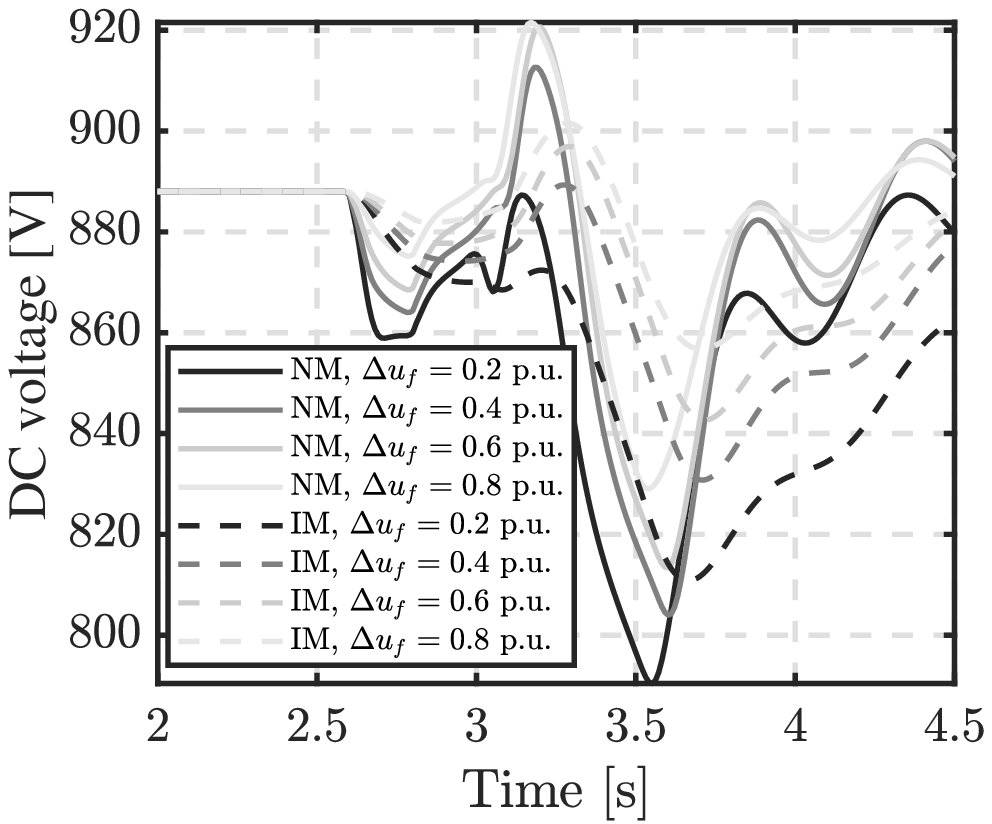}%
\label{fig:lvrt-scdip}}
\caption{DC voltage of a SC bank for a): different fault duration $\tau_{f}$ for a 37\% voltage dip; b): different sizes of the voltage dip $\Delta u_f$ for a 200 ms fault}
\label{fig:lvrt-disturbance}
\end{figure}

The impact of different faults and durations have also been analyzed. Fig. \ref{fig:lvrt-sctime} shows the difference in model performance for different duration (100, 200 and 300 ms) of a fault resulting in a 37\% voltage dip at Bus 06. The initial SoC of the SC is 76--78\%. It can be seen that the nonlinear model will have a somewhat higher voltage spike and a more oscillatory behaviour after the fault clears due to the ESR. Practically, this means that the overvoltage protection may be triggered sooner when a nonlinear model is used compared to an ideal model. Again, the difference between the IM and NM is in tens of V and will not play a significant role in grid dynamics. The longer the fault duration, the bigger the voltage spike and the bigger the difference between models.

The size of the voltage dip (20\% to 80\%) has a similar impact (Fig. \ref{fig:lvrt-scdip}). Ideal model will have smaller and smoother transients than the nonlinear mode due to the lack of ESR, but the observed differences will not significantly impact the grid dynamics. Same as before, depending on the operating point of the SC, the undervoltage or overvoltage protection may be triggered sooner when the nonlinear model is used. 

In summary, the ideal model representation of SC is adequate for transient stability simulations. The difference in internal dynamics behavior is small enough that it should not have any impact on the grid results.
\section{Conclusion}\label{sec:conclusion}
In this paper, an accurate supercapacitor bank model and associated control system has been presented for the use in power system dynamics simulations. Starting from the most detailed RC model of a supercapacitor cell, the model has been gradually reduced until arriving to the simplest representation which adequately describes the supercapacitor dynamics, confirmed by simulation experiments. The proposed model is described with only 4 parameters which are easily obtained from manufacturer's data sheet: capacitance at zero voltage, voltage-dependent capacitance, DC resistance and high-frequency resistance. The performance of the presented model compared to an ideal model has been tested in an IEEE 14-bus test system in frequency control and LVRT scenarios.

For frequency control, the ideal model does not always represent the nonlinear model adequately depending on the initial supercapacitor voltage and disturbance size. For an underfrequency event, a fully to partially charged supercapacitor may be adequately represented by an ideal model in terms of system frequency response, but nearing the minimum voltage limit the ideal model may yield overly optimistic or pessimistic results (frequency nadir difference of over 0.1 Hz can be observed depending on the ideal capacitor capacitance value). Similar behavior is observed for an overfrequency event. Generally, equivalent series resistance and parallel RC groups of the first branch reduce the efficiency of the supercapacitor, while voltage-dependent capacitance changes the amount of stored energy during charging/discharging and influences the charge/discharge rate. The observed mean absolute relative error in discharge time between ideal and nonlinear model ranges from 9\% to 16\% for a 10\% variable capacitance and between 10\% and 25\% for 40\% variable capacitance, while the maximum observed absolute relative error in discharge time can go up to 27\% for 10\% variable capacitance and 43\% for 40\% variable capacitance. The best ideal model for most observed cases for both types of presented control schemes is the ideal model with capacitance set between minimum and average supercapacitor capacitance, with the ideal model with average capacitance being usually more accurate for high initial SoC. Losses and nonlinear voltage dynamics complicate representing a real supercapacitor with an ideal model for all operating points.

For low-voltage ride through, the impact of modelling is not significant and the ideal model will be adequate, although the undervoltage and overvoltage protective circuits may be triggered sooner for the nonlinear model.

\appendix
$n_s$ = 370, $n_p = 400$, bank rated power: $100$ MW, $C_{max} \approx 1000$ F, $R_{dc} = 0.5\text{ m}\Omega$, $R_{s} = 0.25\text{ m}\Omega$, $I_{ch}^{\text{max}}/ I_{dch}^{\text{max}} = \pm 615 A$, $U_{ch}^{\text{max}} = 2.71$ V, $U_{ch}^{\text{start}} = 2.4$ V, $U_{dch}^{\text{min}} = 1.1$ V, $U_{dch}^{\text{start}} = 1.4$ V, $\tau_c = 50$ ms, $K_i = 150$ p.u., $K_d = 100$ p.u., $\tau_w^i = 1$ s, $\tau_w^d = 30$ s, $K_p^d = K_p^q = 1$ p.u., $K_i^d = K_i^q = 100$ p.u.

\bibliographystyle{IEEEtran}
\bibliography{IEEEabrv,references}
\vskip -2\baselineskip plus -1fil
\begin{IEEEbiographynophoto}{Matej Krpan}
(S'17) received the B.Eng. and M.Eng. degrees in electrical engineering from University of Zagreb in 2014 and 2016, respectively. He is currently pursuing the PhD degree at the University of Zagreb Faculty of electrical engineering and computing, Zagreb, Croatia.
\end{IEEEbiographynophoto}
\vskip -2.5\baselineskip plus -1fil
\begin{IEEEbiographynophoto}{Igor Kuzle}
(S'94--M'97--SM'04) is a Full Professor at the Department of Energy and Power Systems of the University of Zagreb Faculty of electrical engineering and computing, Zagreb, Croatia.
\end{IEEEbiographynophoto}
\vskip -2.5\baselineskip plus -1fil
\begin{IEEEbiographynophoto}{Ana Radovanović}
(S'18) received the B.Sc. and M.Sc. degrees in electrical engineering and computer science from the University of Belgrade, Belgrade, Serbia, in 2013 and 2014, respectively. She is currently pursuing the Ph.D. degree at the Department of Electrical and Electronic Engineering, the University of Manchester, Manchester, U.K.
\end{IEEEbiographynophoto}
\vskip -2.5\baselineskip plus -1fil
\begin{IEEEbiographynophoto}{Jovica V. Milanović}
(M'95–SM'98–F'10) is currently a Professor of Electrical Power Engineering and a Deputy Head of Department of Electrical and Electronic Engineering at the University of Manchester, Manchester, U.K. 
\end{IEEEbiographynophoto}






\end{document}